\documentclass[floatfix,aps,prd,twocolumn,letterpaper,lengthcheck,superscriptaddress,showpacs,amssymb,amsmath,amsfonts,nofootinbib,altaffilletter,nopreprintnumbers,showpacs,longbibliography]{revtex4-1}

\usepackage[utf8]{inputenc}
\usepackage{xspace}
\usepackage{acronym}
\usepackage{booktabs}
\usepackage{amsmath,amssymb,graphicx}
\usepackage{color}
\usepackage[colorlinks, linkcolor=blue, pdfborder={0 0 0}, breaklinks=true]{hyperref}
\usepackage[table]{xcolor}
\definecolor{lightgray}{gray}{0.9} % table alternating line colors
\usepackage{tabularx}
\usepackage{multirow}
\usepackage{bigstrut}
\usepackage{braket}
\usepackage{makecell}
\usepackage{mathtools}
\usepackage{lineno}
\usepackage{aas_macros}

\DeclareUnicodeCharacter{2009}{\,}

\newcommand{\norm}[1]{\left\lVert#1\right\rVert}

\begin{document}
	
%\linenumbers

\title{Null-stream-based Bayesian Unmodeled Framework to Probe Generic Gravitational-wave Polarizations}

\author{Isaac C. F.~Wong}
\email{cfwong@link.cuhk.edu.hk}
\affiliation{Department of Physics, The Chinese University of Hong Kong, Shatin, N.T., Hong Kong}
\author{Peter T. H.~Pang}
\email{thopang@nikhef.nl}
\affiliation{Nikhef -- National Institute for Subatomic Physics, Science Park, 1098 XG Amsterdam, The Netherlands}
\affiliation{Institute for Gravitational and Subatomic Physics (GRASP), Utrecht University, Princetonplein 1, 3584 CC Utrecht, The Netherlands}
\author{Rico K. L.~Lo}
\email{kllo@caltech.edu}
\affiliation{LIGO, California Institute of Technology, Pasadena, California 91125, USA}
\author{Tjonnie G. F.~Li}
\affiliation{Department of Physics, The Chinese University of Hong Kong, Shatin, N.T., Hong Kong}
\affiliation{Institute for Theoretical Physics, KU Leuven, Celestijnenlaan 200D, B-3001 Leuven, Belgium}
\affiliation{Department of Electrical Engineering (ESAT), KU Leuven, Kasteelpark Arenberg 10, B-3001 Leuven, Belgium}
\author{Chris Van Den Broeck}
\affiliation{Nikhef -- National Institute for Subatomic Physics, Science Park, 1098 XG Amsterdam, The Netherlands}
\affiliation{Institute for Gravitational and Subatomic Physics (GRASP), Utrecht University, Princetonplein 1, 3584 CC Utrecht, The Netherlands}

\begin{abstract}
	We present a null-stream-based Bayesian unmodeled framework to probe generic gravitational-wave polarizations. Generic metric theories allow six gravitational-wave polarization states, but general relativity only permits the existence of two of them namely the tensorial polarizations. The strain signal measured by an interferometer is a linear combination of the polarization modes and such a linear combination depends on the geometry of the detector and the source location. The detector network of Advanced LIGO and Advanced Virgo allows us to measure different linear combinations of the polarization modes and therefore we can constrain the polarization content by analyzing how the polarization modes are linearly combined. We propose the basis formulation to construct a null stream along the polarization basis modes without requiring modeling the basis explicitly. We conduct a mock data study and we show that the framework is capable of probing pure and mixed polarizations in the Advanced LIGO-Advanced Virgo 3-detector network without knowing the sky location of the source from electromagnetic counterparts. We also discuss the effect of the presence of the uncaptured orthogonal polarization component in the framework, and we propose using the plug-in method to test the existence of the orthogonal polarizations.
\end{abstract}

\maketitle

\section{introduction}

Since the first gravitational wave (GW) detection \cite{PhysRevLett.116.061102} in 2015, Advanced LIGO \cite{2015} and Advanced Virgo \cite{Acernese_2014} have detected dozens of compact binary coalescence (CBC) events in the first, second and third observing runs \cite{PhysRevX.9.031040,abbott2020gwtc2}. The GW detections allow us to test general relativity (GR) in the strong-field and dynamical regime \cite{PhysRevD.100.104036,theligoscientificcollaboration2020tests}.

One property of GWs predicted by GR is that they are described by only two polarization modes namely the plus polarization and the cross polarization or collectively the tensorial polarizations. Generic metric theories allow six GW polarization modes \cite{PhysRevD.8.3308} which are two tensorial polarization modes, two vectorial polarization modes, and two scalar polarization modes. Non-tensorial polarization modes exist in many modified gravity theories. For example, the Brans-Dicke theory \cite{PhysRev.124.925} predicts a scalar breathing polarization mode in addition to the two tensorial polarization modes. The bimetric theory proposed by Rosen \cite{PhysRevD.3.2317,ROSEN1974455} predicts the existence of all six polarization modes. Different polarization modes stretch and squeeze the space differently, and the strain measured by an interferometer is a linear combination of the polarization modes. The polarization modes are linearly combined in a way that depends on the geometry of the interferometer and the source location. A network of non-coaligned interferometers allows us to measure different linear combinations of the polarization modes in each detector and therefore allows us to constrain the polarization content of the signal. 

Different methods have been proposed to constrain the polarization content of GWs. At the time of writing the coincident detections are only up to three detectors, so the analyses so far only focus on pure polarizations i.e.\ comparing the likelihood of the signal being purely tensorial, purely vectorial, and pure scalar since at least $M+1$ non-coaligned detectors are needed to resolve $M$ non-degenerate polarization modes \cite{PhysRevD.86.022004}. To date, the proposed tests could be categorized into two groups which are heuristic tests and model-independent tests. The heuristic test in Refs.~\cite{PhysRevLett.119.141101,Abbott_2019_tgr} is performed by replacing the tensorial beam pattern function with the non-tensorial beam pattern function while keeping the GR waveform template. The model-independent tests include the sine-Gaussian analysis using \texttt{BayesWave} \cite{Cornish_2015} performed on GW150914 \cite{PhysRevLett.116.221101}, and the null-stream-based analysis \cite{theligoscientificcollaboration2020tests} performed on GWTC-2 \cite{abbott2020gwtc2}.

In our previous work \cite{Pang_2020}, we discussed null-stream-based frequentist methods to probe mixed polarizations with the source location informed by electromagnetic (EM) counterparts. The methods do not rely on any waveform models, and we could therefore test the polarization content model-agnostically to confirm or rule out a specific group of modified gravity theories. However, there are two disadvantages of the methods. First, it requires the source location to be known that could be informed by EM counterparts, but to date, GW170817 is the only GW event that has a confident association with a gamma-ray burst \cite{Abbott_2017} and this indicates the rarity of joint GW-EM observations. Second, the methods could only detect the existence of non-tensorial polarizations but are not capable of inferring which non-tensorial components are more likely to present. In this work, we propose a generalized null-stream-based Bayesian framework to infer generic GW polarizations including pure and mixture polarizations without requiring the a priori knowledge of source location. The Bayesian framework allows us to compare the marginal likelihood between different polarization hypotheses in contrast to the frequentist methods that we previously proposed.

The paper is structured as follows. In Sec.~\ref{sec:method}, we review the observation model of GW and null stream, and we discuss the basis formulation and develop the Bayesian framework to probe GW polarizations. In Sec.~\ref{sec:results}, we present the results of a mock data study with non-GR injections and show that our framework is capable of probing GW polarizations without requiring the a priori knowledge of the source location. We also discuss the effect of the presence of an uncaptured orthogonal polarization component in the analysis, and we present the plug-in method to test the existence of an orthogonal polarization component in the signal. The conclusions are summarized in Sec.~\ref{sec:conclusion}.
\section{Methodology}
\label{sec:method}

In this section, we describe the GW observation model, review the null stream, and present the Bayesian null stream formulation to probe GW polarizations.

\subsection{Observation model}
\label{sec:obs_model}

\begin{figure}
\begin{center}
	\includegraphics[width=0.8\linewidth]{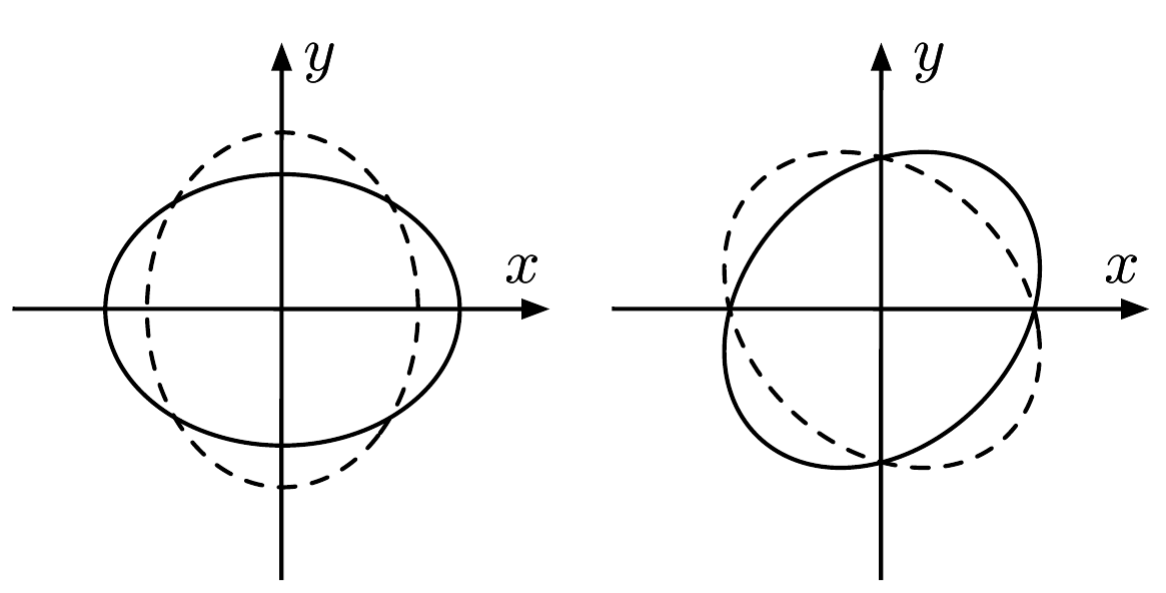}
	\includegraphics[width=0.8\linewidth]{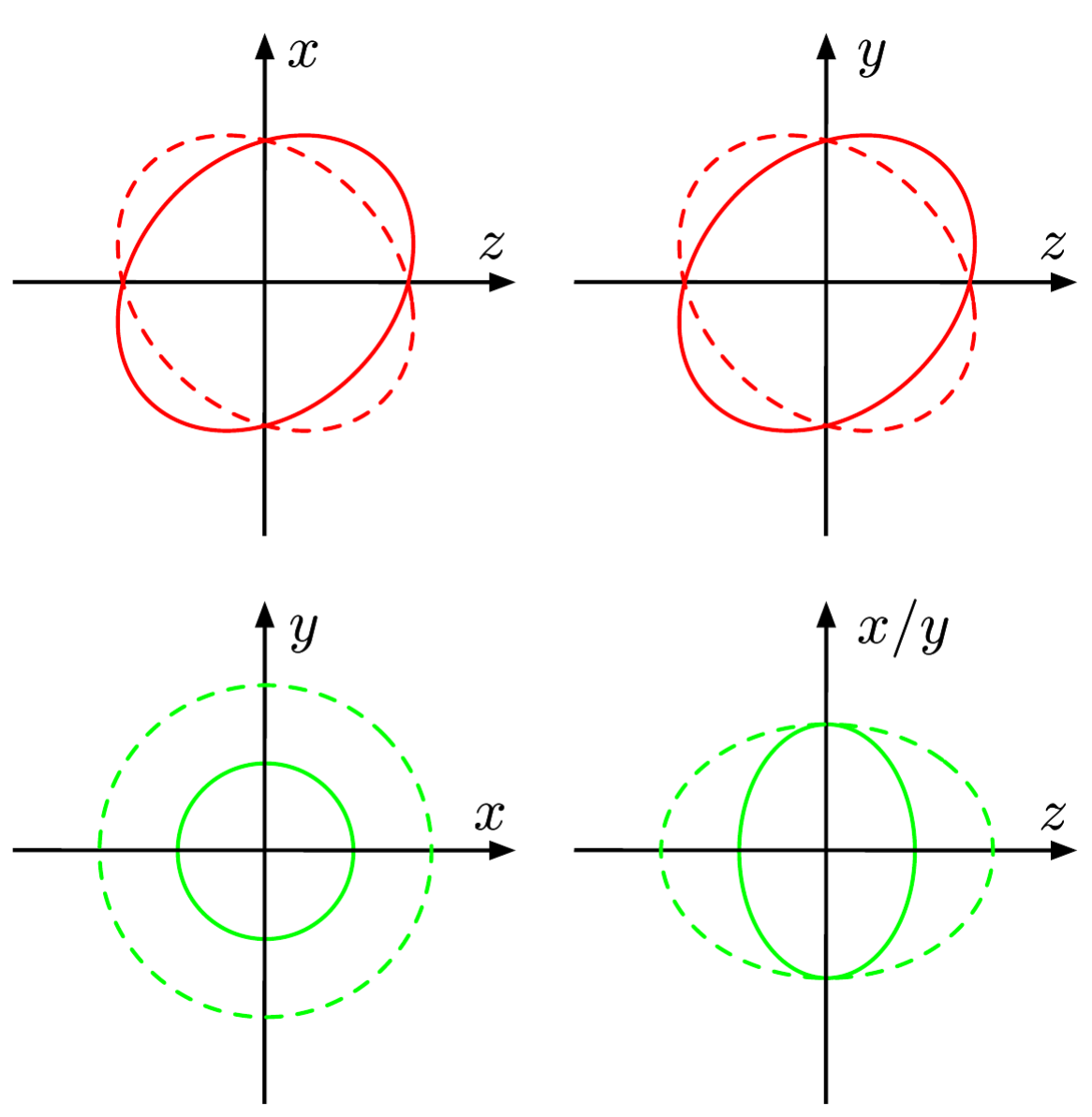}
\end{center}
\caption[Effects of gravitational waves with non-general-relativity polarization modes on a ring of test mass]{The effect on a ring of free-falling test particles of a GW in tensor ``$+$" mode (upper left), tensor ``$\times$" mode (upper right), vector ``X"  mode (middle left), vector ``Y" mode (middle right), scalar breathing mode (lower left) and scalar longitudinal mode (lower right). In each case the wave is traveling in the $z$-direction. The solid and dotted lines are the states of the ring with a phase difference of $\pi$.}
\label{fig:pol_demo}
\end{figure}

The additive noise observation model of GW with all possible polarization modes in a $D$-detector network writes
\begin{equation}
\boldsymbol{d}(t;\Delta\boldsymbol{t}) = \boldsymbol{F}(\alpha,\delta,\psi, t)\boldsymbol{h}(t)+\boldsymbol{n}(t;\Delta\boldsymbol{t})
\label{eq:obs_model_0}
\end{equation}
where
\begin{equation}
\boldsymbol{d}(t;\Delta\boldsymbol{t})=
\begin{bmatrix}
d_{1}(t+\Delta t_{1}) \\
d_{2}(t+\Delta t_{2}) \\
\vdots \\
d_{D}(t+\Delta t_{D})
\end{bmatrix}
\end{equation}
is the observed strain outputs at shifted times,
\begin{equation}
\begin{split}
\boldsymbol{F}(\alpha, \delta, \psi, t)
&=\begin{bmatrix}
\boldsymbol{f}_{+} & \boldsymbol{f}_{\times} & \boldsymbol{f}_{b} & \boldsymbol{f}_{l} & \boldsymbol{f}_{x} & \boldsymbol{f}_{y}
\end{bmatrix}\\
&=\begin{bmatrix}
F_{1}^{+} & F_{1}^{\times} & F_{1}^{b} & F_{1}^{l} & F_{1}^{x} & F_{1}^{y} \\
F_{2}^{+} & F_{2}^{\times} & F_{2}^{b} & F_{2}^{l} & F_{2}^{x} & F_{2}^{y} \\
\vdots & \vdots  & \vdots & \vdots & \vdots & \vdots \\
F_{D}^{+} & F_{D}^{\times} & F_{D}^{b} & F_{D}^{l} & F_{D}^{x} & F_{D}^{y}
\end{bmatrix}
\end{split}
\end{equation}
is the beam pattern matrix,
\begin{equation}
\boldsymbol{h}(t)=
\begin{bmatrix}
h_{+}(t) &
h_{\times}(t) &
h_{b}(t) &
h_{l}(t) &
h_{x}(t) &
h_{y}(t)
\end{bmatrix}^{T}
\end{equation}
are the polarization modes where $T$ denotes transpose and the plus mode, cross mode, scalar breathing mode, scalar longitudinal mode, vector x mode and vector y mode are labeled with the subscripts $+$, $\times$, $b$, $l$, $x$ and $y$ respectively,
\begin{equation}
\boldsymbol{n}(t;\Delta\boldsymbol{t})=
\begin{bmatrix}
n_{1}(t+\Delta t_{1}) \\
n_{2}(t+\Delta t_{2}) \\
\vdots \\
n_{D}(t+\Delta t_{D})
\end{bmatrix}
\end{equation}
is the detector noise at shifted times, $\alpha$, $\delta$ are the right ascension and declination of the source location respectively, $\psi$ is the polarization angle, $\Delta\boldsymbol{t}=\{\Delta t_{1}, \Delta t_{2}, ..., \Delta t_{D}\}$ where
\begin{equation}
\Delta t_{j} = -\frac{\boldsymbol{r}_{j}\cdot\hat{\boldsymbol{N}}}{c}
\end{equation}
is the time delay of the signal arrival at each detector with reference to the Earth center, $\boldsymbol{r}_{j}$ is the coordinates of the $j^{\text{th}}$ detector in the Earth-centered coordinate system,
\begin{equation}
\hat{\boldsymbol{N}}=
\begin{bmatrix}
\cos(\delta)\cos(t_{\text{gmst}}-\alpha) \\ -\cos(\delta)\sin(t_{\text{gmst}}-\alpha) \\
\sin(\delta)
\end{bmatrix}
\end{equation}
where $t_{\text{gmst}}$ is the Greenwich mean sidereal time, and $c$ is the speed of gravity. For a two-arm interferometer, the beam pattern functions take the following forms when the two arms are directed to the x-axis and y-axis respectively \cite{Nishizawa_2009}:
\begin{equation}
\begin{split}
F_{+}(\theta,\phi,\psi)=&\frac{1}{2}(1+\cos^{2}\theta)\cos{2\phi}\cos{2\psi}\\
&-\cos\theta\sin{2\phi}\sin{2\psi}
\end{split}
\end{equation}
\begin{equation}
\begin{split}
F_{\times}(\theta,\phi,\psi)=&-\frac{1}{2}(1+\cos^{2}\theta)\cos{2\phi}\sin{2\psi}\\
&-\cos{\theta}\sin{2\phi}\cos{2\psi}
\end{split}
\end{equation}
\begin{equation}
F_{x}(\theta,\phi,\psi)=\sin{\theta}(\cos{\theta}\cos{2\phi}\cos{\psi}-\sin{2\phi}\sin{\psi})
\end{equation}
\begin{equation}
F_{y}(\theta,\phi,\psi)=-\sin{\theta}(\cos\theta\cos{2\phi}\sin{\psi}+\sin{2\phi}\cos{\psi})
\end{equation}
\begin{equation}
\label{eq:F_b}
F_{b}(\theta,\phi)=-\frac{1}{2}\sin^{2}\theta\cos{2\phi}
\end{equation}
\begin{equation}
\label{eq:F_l}
F_{l}(\theta,\phi)=\frac{1}{\sqrt{2}}\sin^{2}\theta\cos{2\phi}
\end{equation}
where $\theta$ and $\phi$ are the polar angle and the azimuthal angle of the source location in the Earth-centered coordinate system respectively, and $\psi$ is the polarization angle. Fig.~\ref{fig:pol_demo} demonstrates the effect of different polarization modes on a ring of free-falling test particles.

In frequency domain, the observation model writes
\begin{equation}
\tilde{\boldsymbol{d}}[k;\Delta\boldsymbol{t}] = \boldsymbol{F}(\alpha,\delta,\psi,t_{\text{event}})\tilde{\boldsymbol{h}}[k]+\tilde{\boldsymbol{n}}[k;\Delta\boldsymbol{t}]
\end{equation}
where
\begin{equation}
\tilde{\boldsymbol{d}}[k;\Delta\boldsymbol{t}]=
\begin{bmatrix}
\tilde{d}_{1}[k]e^{i2\pi k\Delta t_{1}/T} \\
\tilde{d}_{2}[k]e^{i2\pi k\Delta t_{2}/T} \\
\vdots \\
\tilde{d}_{D}[k]e^{i2\pi k\Delta t_{D}/T}
\end{bmatrix}
\end{equation}
is the Fourier transform (see Appendix \ref{app:DFT}) of the time-shifted strain outputs, $T$ is the duration of data in second, $\boldsymbol{F}(\alpha,\delta,\psi,t)\approx \boldsymbol{F}(\alpha,\delta,\psi,t_{\text{event}})$ is approximated to be constant over the data segment which is valid for a transient signal and is evaluated at the event time $t_{\text{event}}$,
\begin{equation}
\tilde{\boldsymbol{h}}[k]=
\begin{bmatrix}
\tilde{h}_{+}[k] & \tilde{h}_{\times}[k] & \tilde{h}_{b}[k] & 	\tilde{h}_{l}[k] & \tilde{h}_{x}[k] & \tilde{h}_{y}[k]
\end{bmatrix}^{T}
\end{equation}
is the Fourier transform of the polarization modes,
and
\begin{equation}
\tilde{\boldsymbol{n}}[k;\Delta\boldsymbol{t}]=
\begin{bmatrix}
\tilde{n}_{1}[k]e^{i2\pi k\Delta t_{1}/T} \\
\tilde{n}_{2}[k]e^{i2\pi k\Delta t_{2}/T} \\
\vdots \\
\tilde{n}_{D}[k]e^{i2\pi k\Delta t_{D}/T}
\end{bmatrix}
\end{equation}
is the Fourier transform of the time-shifted noise in each detector. Assume the detector noise follows a stationary Gaussian distribution and the noise is independent between detectors, we have
\begin{equation}
\braket{\tilde{n}_{j'}^{*}[k']\tilde{n}_{j}[k]}=\frac{1}{2\Delta f}S_{j}[k]\delta_{k'k}\delta_{j'j}
\label{eq:psd}
\end{equation}
where $\braket{\cdot}$ denotes expectation over the noise realizations, $S_{j}[k]$ is the one-sided power spectral density (PSD) of the noise in detector $j$, $\delta_{jk}$ denotes Kronecker delta, and $\Delta f$ is the frequency resolution of the discrete Fourier transform.

For brevity, in the following sections we write the time-shifted data as the time-shift operator $\boldsymbol{\mathcal{T}}\left(\cdot;\Delta\boldsymbol{t}\right):\mathbb{C}^{D\times K}\rightarrow\mathbb{C}^{D\times K}$ applied on the unshifted data defined by
\begin{equation}
	\boldsymbol{\mathcal{T}}(\tilde{\boldsymbol{x}};\Delta\boldsymbol{t}) = \tilde{\boldsymbol{x}}\odot\mathcal{S}(\Delta\boldsymbol{t})
\end{equation}
where
\begin{equation}
	\tilde{\boldsymbol{x}}=\begin{bmatrix}
	\tilde{x}_{1}[1] & \tilde{x}_{1}[2] & \ldots & \tilde{x}_{1}[K] \\
	\tilde{x}_{2}[1] & \tilde{x}_{2}[2] & \ldots & \tilde{x}_{2}[K] \\
	\vdots & \vdots & \ddots & \vdots \\
	\tilde{x}_{D}[1] & \tilde{x}_{D}[2] & \ldots & \tilde{x}_{D}[K] 
	\end{bmatrix}
\end{equation}
is the unshifted data, $(\mathcal{S}(\Delta\boldsymbol{t}))_{jk}=e^{i2\pi (k-1)\Delta t_{j}/T}$ where $j=1,2,...,D$ and $k=1,2,...,K$ or more explicitly
\begin{equation}
	\begin{split}
	&\mathcal{S}(\Delta\boldsymbol{t})\\
	&=
	\begin{bmatrix}
	e^{i2\pi (0)\Delta t_{1}/T} & e^{i2\pi (1)\Delta t_{1}/T} & \ldots & e^{i2\pi (K-1)\Delta t_{1}/T} \\
	e^{i2\pi (0)\Delta t_{2}/T} & e^{i2\pi (1)\Delta t_{2}/T} & \ldots & e^{i2\pi (K-1)\Delta t_{2}/T} \\
	\vdots & \vdots & \ddots & \vdots \\
	e^{i2\pi (0)\Delta t_{D}/T} & e^{i2\pi (1)\Delta t_{D}/T} & \ldots & e^{i2\pi (K-1)\Delta t_{D}/T}
	\end{bmatrix}
	\end{split}\,,
\end{equation}
$\odot$ denotes the elementwise product, and $K$ is the number of frequency bins.
\subsection{Null Stream}
\label{sec:null_stream}

\begin{figure}
	\includegraphics[width=\linewidth]{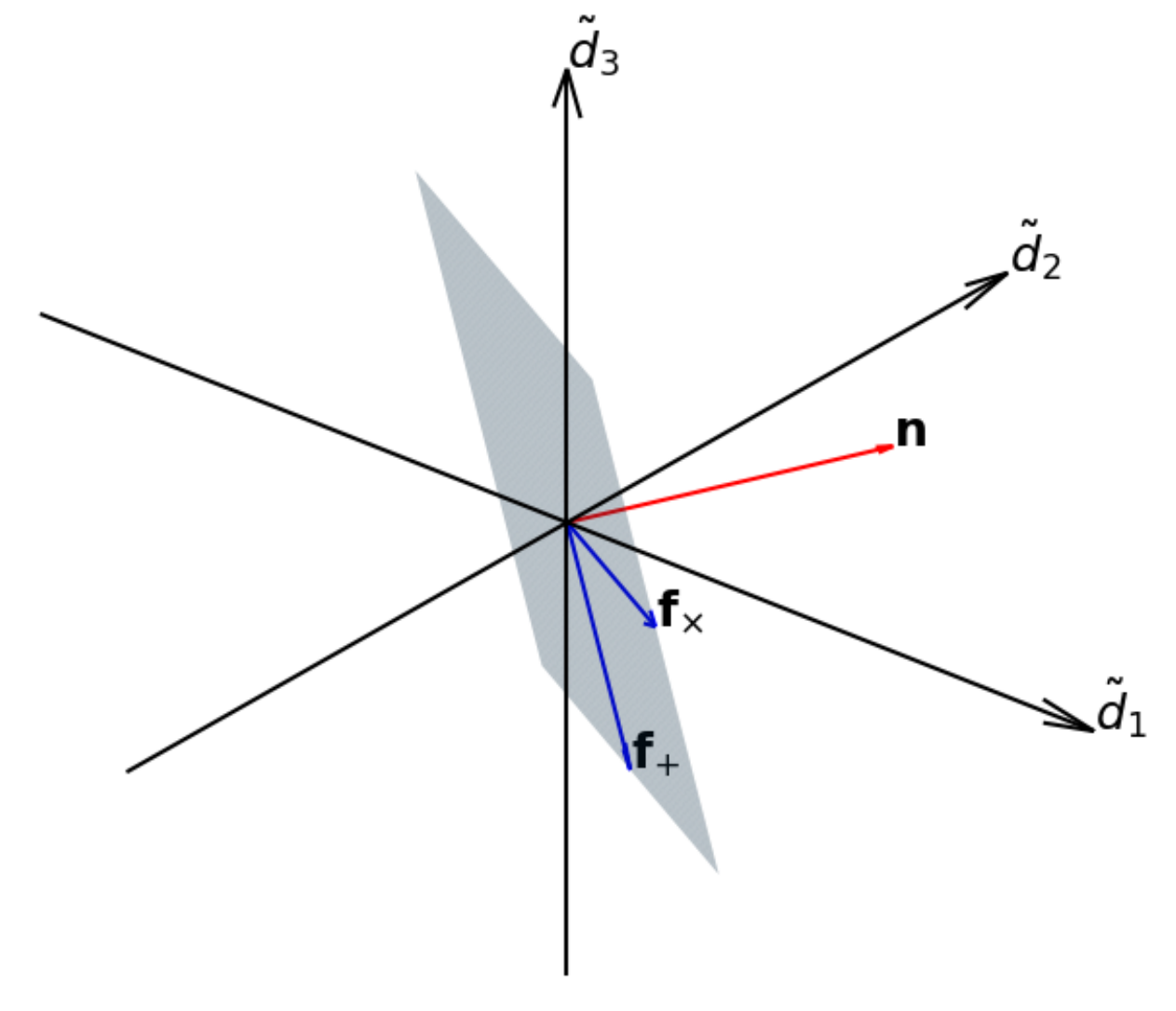}
	\caption{The axes $\tilde{d}_{1}$, $\tilde{d}_{2}$ and $\tilde{d}_{3}$ represent the measurement of the strain output in each detector. The hyperplane spanned by $\boldsymbol{f}_{+}$ and $\boldsymbol{f}_{\times}$ defines all possible strain measurement of a tensorial signal in each detector. The vector $\boldsymbol{n}$ is orthogonal to $\boldsymbol{f}_{+}$ and $\boldsymbol{f}_{\times}$ and it spans the null space of $\boldsymbol{f}_{+}$ and $\boldsymbol{f}_{\times}$. There is no strain measurement of any tensorial signal along $\boldsymbol{n}$.}
	\label{fig:hyperplane}
\end{figure}

\textit{Null stream} \cite{PhysRevD.40.3884} is a specific linear combination of strain outputs from a network of detectors such that the linear combination only contains noise regardless of the GW waveform. Let us consider a tensorial signal as follows
\begin{align}
	\tilde{\boldsymbol{s}}(f) = 
	\begin{bmatrix}
		\boldsymbol{f}_{+} & \boldsymbol{f}_{\times}
	\end{bmatrix}
	\begin{bmatrix}
		\tilde{h}_{+}(f) \\
		\tilde{h}_{\times}(f)
	\end{bmatrix}\,,
\end{align}
the signal $\tilde{\boldsymbol{s}}(f)$ could be regarded as living in the subspace spanned by $\boldsymbol{f}_{+}$ and $\boldsymbol{f}_{\times}$ as shown in Fig.~\ref{fig:hyperplane}. The hyperplane spanned by $\boldsymbol{f}_{+}$ and $\boldsymbol{f}_{\times}$ in the figure defines all possible strain measurement of a tensorial signal in each detector. We could project away the hyperplane to remove any tensorial signal regardless of the waveform of $\tilde{h}_{+}(f)$ and $\tilde{h}_{\times}(f)$. The residual remains in the null space of $\boldsymbol{f}_{+}$ and $\boldsymbol{f}_{\times}$ is the so-called null stream which contains noise only.

One application of null stream is to distinguish between coincident GW transients and coincident noise glitches. If the observed coherent excess power is a genuine GW transient, one should obtain a null stream consistent with noise. For noise glitches, in general, the resultant null stream would still contain excess power. Null stream is therefore used in conducting vetos \cite{Wen_2005,Ajith_2006,Chatterji_2006} and GW searches associated with gamma-ray bursts and other astrophysical triggers \cite{Sutton_2010}. In this section, we review the method to construct a null stream. The matrix representation follows from Ref.~\cite{Sutton_2010}. The only difference is that we include the time-shift operation explicitly in the construction of null stream since we do not know the source location a priori in contrast to the follow-up analysis of gamma-ray bursts in Ref.~\cite{Sutton_2010}.

Suppose we observe a signal at time $t_{\text{event}}$, the frequency domain observation model writes
\begin{equation}
	\label{eq:obs_model}
	\tilde{\boldsymbol{d}}[k]=\boldsymbol{\mathcal{T}}\left(\boldsymbol{F}(\alpha,\delta,\psi,t_{\text{event}})\tilde{\boldsymbol{h}};-\Delta\boldsymbol{t}\right)[k]+\tilde{\boldsymbol{n}}[k]\,.
\end{equation}
Before performing the null projection, we first whiten the data since this would make it easier to handle the statistics of the residual noise. Dividing both sides of Eq.~(\ref{eq:obs_model}) by $\sqrt{\frac{1}{2\Delta f}S[k]}$, we have
\begin{equation}
	\tilde{\boldsymbol{d}}_{w}[k]=\boldsymbol{\mathcal{T}}\left(\boldsymbol{F}_{w}(\alpha,\delta,\psi,t_{\text{event}})\tilde{\boldsymbol{h}};-\Delta \boldsymbol{t}\right)[k]+\tilde{\boldsymbol{n}}_{w}[k]
	\label{eq:d_w}
\end{equation}
where
\begin{equation}
	\tilde{d}_{w,j}[k]=\frac{\tilde{d}_{j}[k]}{\sqrt{\frac{1}{2\Delta f}S_{j}[k]}}\,,
\end{equation}
\begin{equation}
    \label{eq:Fw}
	F_{w,j}[k]=\frac{F_{j}}{\sqrt{\frac{1}{2\Delta f}S_{j}[k]}}
\end{equation}
and
\begin{equation}
	\tilde{n}_{w,j}[k]=\frac{\tilde{n}_{j}[k]}{\sqrt{\frac{1}{2\Delta f}S_{j}[k]}}
\end{equation}
where $F_{j}$ represents the $j^{\text{th}}$ row of $\boldsymbol{F}$, $\boldsymbol{F}_{w}\in\mathbb{C}^{D\times M\times K}$ is the noise-weighed beam pattern matrix, and $M$ is the number of column vectors in $\boldsymbol{F}$. The construction of null stream consists of two steps: 1) time-shifting the data with the time delay implied by the source location and event time i.e.\ $\Delta\boldsymbol{t}=\Delta\boldsymbol{t}(\alpha,\delta,t_{\text{event}})$, and 2) performing the null projection. Denote the null operator as $\boldsymbol{P}: \mathbb{C}^{D\times D\times K}\rightarrow\mathbb{C}^{D\times D \times K}$ defined by
\begin{equation}
	\label{eq:P_def}
	\boldsymbol{P}\left({}\cdot{};\alpha,\delta,\psi,t_{\text{event}}\right)=(\boldsymbol{I}-\boldsymbol{F}_{w}(\boldsymbol{F}_{w}^{\dagger}\boldsymbol{F}_{w})^{-1}\boldsymbol{F}_{w}^{\dagger})\boldsymbol{\mathcal{T}}\left(\cdot;\Delta\boldsymbol{t}\right)
\end{equation}
where $\boldsymbol{I}\in\mathbb{C}^{D\times D\times K}$ is a stack of $K$ $D\times D$ identity matrices and $\dagger$ denotes conjugate transpose. The matrix operation in Eq.~\eqref{eq:P_def} is performed frequency-wise. One should notice that the null operator is completely determined by the constituent $\boldsymbol{f}_{m}$ of $\boldsymbol{F}$ and the set of parameters $\{\alpha,\delta,\psi,t_{\text{event}}\}$ is independent of the waveform $\tilde{\boldsymbol{h}}$. When the set of parameters are correctly specified, one could verify that the null operator projects away $\tilde{\boldsymbol{h}}$ without requiring any knowledge of $\tilde{\boldsymbol{h}}$ itself, and the residual $\tilde{\boldsymbol{z}}\in\mathbb{C}^{D\times K}$ is
\begin{equation}
\label{eq:residual_start}
\begin{split}    
\tilde{\boldsymbol{z}}&\coloneqq\boldsymbol{P}(\tilde{\boldsymbol{d}}_{w};\alpha,\delta,\psi,t_{\text{event}}) \\
&=(\boldsymbol{I}-\boldsymbol{F}_{w}(\boldsymbol{F}_{w}^{\dagger}\boldsymbol{F}_{w})^{-1}\boldsymbol{F}_{w}^{\dagger})\boldsymbol{\mathcal{T}}\left(\tilde{\boldsymbol{d}}_{w};\Delta\boldsymbol{t}\right) \\
&=(\boldsymbol{I}-\boldsymbol{F}_{w}(\boldsymbol{F}_{w}^{\dagger}\boldsymbol{F}_{w})^{-1}\boldsymbol{F}_{w}^{\dagger})\left(\boldsymbol{F}_{w}\tilde{\boldsymbol{h}}+\boldsymbol{\mathcal{T}}(\tilde{\boldsymbol{n}};\Delta\boldsymbol{t})\right)\\
&=(\boldsymbol{I}-\boldsymbol{F}_{w}(\boldsymbol{F}_{w}^{\dagger}\boldsymbol{F}_{w})^{-1}\boldsymbol{F}_{w}^{\dagger})\boldsymbol{\mathcal{T}}\left(\tilde{\boldsymbol{n}}_{w};\Delta\boldsymbol{t}\right)
\end{split}
\end{equation}
which is the residual noise in the lower dimensional space. To understand the reduced dimensionality, one should be reminded that null projection removes the data on the hyperplane spanned by the beam pattern vectors $\boldsymbol{f}_{m}$, and we may call the hyperplane as the GW space. Only the data in the null space that is orthogonal to the GW space survive and it is the so-called null stream. One could perform a rotation $\mathcal{R}$ to rotate the residuals to align along the principal axes to visualize the reduced dimensionality (see Appendix \ref{app:reduced_dim}). Although we write the null operator in Eq.~\eqref{eq:P_def} as a function of the polarization angle $\psi$, the null operator is in fact independent of the polarization angle if we consider non-degenerate polarization modes (see Appendix \ref{app:pol_angle}).
\subsection{Basis formulation}
\label{sec:formulation}

In this section, we present the basis formulation to test GW polarizations. Suppose we would like to compare two polarization hypotheses $\mathcal{H}_{1}$ and $\mathcal{H}_{2}$ which beam pattern matrices are $\boldsymbol{F}_{1}\in\mathbb{R}^{D\times m_{1}}$ and $\boldsymbol{F}_{2}\in\mathbb{R}^{D\times m_{2}}$ respectively where $m_{1}<m_{2}$ without loss of generality. One should notice the resultant null streams have unequal dimensionalities. To have a fair comparison between different models, the residuals have to be of the same dimensionality. Instead of performing the most generic null projection with $\boldsymbol{F}_{2}\in\mathbb{R}^{D\times m_{2}}$, one could perform the null projection to remove a smaller subspace to match the dimensionality of $\boldsymbol{F}_{1}\in\mathbb{R}^{D\times m_{1}}$. We propose the basis formulation to restrict the null projection to remove a smaller subspace constrained by the polarization basis modes. To illustrate the idea, we may consider a scalar-tensorial signal model with the scalar breathing mode in addition to the two tensorial modes
\begin{equation}
	\label{eq:TS_model}
	\tilde{\boldsymbol{s}}[k] = \boldsymbol{f}_{+}\tilde{h}_{+}[k] + \boldsymbol{f}_{\times}\tilde{h}_{\times}[k] + \boldsymbol{f}_{b}\tilde{h}_{b}[k]\,.
\end{equation}
The most generic null projection would remove three dimensions from the data since we have three linearly independent beam pattern vectors. Without loss of generality, we choose the $+$ mode as the basis, and we can always decompose any polarization mode $\tilde{h}_{m}[k]$ into the parallel component and the orthogonal component with respect to the basis i.e.
\begin{equation}
	\tilde{h}_{m}[k] = C_{\parallel}^{m}\tilde{h}_{\parallel}[k] + C_{\perp}^{m}\tilde{h}_{\perp}[k]
\end{equation}
where $\tilde{h}_{\parallel}[k] = \tilde{h}_{+}[k]$ and $C_{\parallel}, C_{\perp}\in\mathbb{C}$. With this formulation, we can rewrite Eq.~\eqref{eq:TS_model} as
\begin{equation}
\begin{split}
	&\tilde{\boldsymbol{s}}[k] \\
	&= (\boldsymbol{f}_{+} + C_{\parallel}^{\times}\boldsymbol{f}_{\times} + C_{\parallel}^{b}\boldsymbol{f}_{b})\tilde{h}_{\parallel}[k] +
	(C_{\perp}^{\times}\boldsymbol{f}_{\times} + C_{\perp}^{b}\boldsymbol{f}_{b})\tilde{h}_{\perp}[k] \\
	&=\boldsymbol{f}_{\parallel}\tilde{h}_{\parallel}[k] + \boldsymbol{f}_{\perp}\tilde{h}_{\perp}[k]
\end{split}
\end{equation}
where
\begin{equation}
\begin{split}
	&\tilde{h}_{+}[k] = \tilde{h}_{\parallel}[k] \\ &\tilde{h}_{\times}[k] = C_{\parallel}^{\times}\tilde{h}_{\parallel}[k] + C_{\perp}^{\times}\tilde{h}_{\perp}[k] \\
	&\tilde{h}_{b}[k] = C_{\parallel}^{b}\tilde{h}_{\parallel}[k] + C_{\perp}^{b}\tilde{h}_{\perp}[k]
\end{split}
\end{equation}
and
\begin{equation}
\begin{split}
	&\boldsymbol{f}_{\parallel} = \boldsymbol{f}_{+} + C_{\parallel}^{\times}\boldsymbol{f}_{\times} + C_{\parallel}^{b}\boldsymbol{f}_{b} \\
	&\boldsymbol{f}_{\perp} = C_{\perp}^{\times}\boldsymbol{f}_{\times} + C_{\perp}^{b}\boldsymbol{f}_{b}\,.
\end{split}
\end{equation}
We could then perform the (partial) null projection to remove the smaller subspace spanned by $\boldsymbol{f}_{\parallel}$ only. Qualitatively, we have reduced the most generic scalar-tensor GW space spanned by $\boldsymbol{f}_{+}$, $\boldsymbol{f}_{\times}$ and $\boldsymbol{f}_{b}$ to the scalar-tensor GW subspace spanned by $\boldsymbol{f}_{\parallel}$ where the polarization modes are described by a single basis. With this formulation, the dimensionality of the GW space is then determined by the number of basis modes in contrast to that in the generic null projection determined by the number of beam pattern vectors. In other words, this formulation allows us to capture the non-tensorial components parallel to the basis mode(s) (in general we can have more than one basis mode). One attractive feature of this formulation is that we only need to choose the basis modes of some polarization types, but we do not need to assume the waveform of the basis modes.

In the basis formulation, we have the freedom to choose the basis mode(s) without explicitly modeling them. Also, we have the freedom to choose the number of basis mode(s). In terms of testing GR, we want to compare the likelihood between the tensor hypothesis and the non-tensor hypotheses. Since there are two polarization modes in tensorial polarizations, we could choose either one or two basis mode(s) to parameterize the tensor hypothesis. The competing hypotheses are then formulated with the same number of basis modes accordingly. For brevity, we may call the one-basis-mode analysis as the $L=1$ analysis, and the two-basis-mode analysis as the $L=2$ analysis where $L$ denotes the number of basis mode(s). In the $L = 2$ analysis of the tensor hypothesis, the signal space contains all possible tensorial polarizations, and the waveforms of the plus and cross polarization can take any independent shape. In the $L = 1$ analysis, an additional structure is imposed on the plus and cross polarizations i.e.\ they are linearly dependent in the Fourier domain, and therefore the signal space is more restricted compared to that in the $L = 2$ analysis.

In general, a higher number of basis modes gives a greater degree of freedom to fit the data but it also reduces the distinguishing power between different polarization hypotheses. The major reason is that with a higher number of basis modes we remove a larger subspace from the data, and therefore the residual has a lower dimensionality. One should be reminded that the polarization subspaces spanned by the beam pattern vectors are in general not orthogonal to each other. For example, removing the tensorial subspace could also remove part of the non-tensorial component (if any) of the signal. The polarization subspaces, in general, have a greater overlap in the $L = 2$ analysis than that in the $L = 1$ analysis, and one should expect the analysis with a higher number of basis modes has a weaker distinguishing power between the polarization hypotheses.

Although the $L=1$ analysis, in general, has a stronger distinguishing power than the $L=2$ analysis, the $L=2$ analysis helps to capture the orthogonal polarization mode(s) that is/are possibly missed in the $L=1$ analysis. In practice, we can always perform both analyses and examine the consistency of the results.

\subsection{Parameterization}

In this section, we present the parameterization of the $L=1$ analysis and the $L=2$ analysis. Each polarization hypothesis is characterized by the beam pattern vector $\boldsymbol{f}_{m}$ involved in the construction of the null operator. To facilitate the mathematical expressions, we may define $\mathbb{F}_{p}$ as the set of labels of polarizations corresponding to the polarization hypothesis $\mathcal{H}_{p}$ as follows
\begin{equation}
\label{eq:pol_set}
\begin{split}
	&\mathbb{F}_{T} = \{+, \times\} \\
	&\mathbb{F}_{V} = \{x, y\} \\
	&\mathbb{F}_{S} = \{b, l\} \\
	&\mathbb{F}_{TV} = \mathbb{F}_{T} \cup \mathbb{F}_{V} \\
	&\mathbb{F}_{TS} = \mathbb{F}_{T} \cup \mathbb{F}_{S} \\
	&\mathbb{F}_{VS} = \mathbb{F}_{V} \cup \mathbb{F}_{S} \\
	&\mathbb{F}_{TVS} = \mathbb{F}_{T} \cup \mathbb{F}_{V} \cup \mathbb{F}_{S}
\end{split}
\end{equation}
where $+$, $\times$, $b$, $l$, $x$ and $y$ denotes the labels for the plus polarization, cross polarization, scalar breathing polarization, scalar longitudinal polarization, vector $x$ polarization, and vector $y$ polarization respectively.

\subsubsection{One-basis-mode ($L=1$) analysis}
\label{sec:L_1}

In the $L = 1$ analysis, the beam pattern matrix $\boldsymbol{F}$ of the polarization hypothesis $\mathcal{H}_{p}$ with the basis mode $\ell$ is defined by
\begin{equation}
\label{eq:L1_pol}
\begin{split}
	&\boldsymbol{F}(\alpha, \delta, \psi, \boldsymbol{\mathcal{A}}, \boldsymbol{\varphi}, t_{\text{event}}) \\
	&=\begin{bmatrix}
		\boldsymbol{f}_{\ell} + \sum\limits_{m\in\mathbb{F}_{p} \setminus \{\ell\} }
		\mathcal{A}_{m}e^{i\varphi_{m}}\boldsymbol{f}_{m}
	\end{bmatrix}\,.
\end{split}
\end{equation}
where $\mathbb{F}_{p} \setminus \{\ell\}$ is the set of labels of polarizations in hypothesis $\mathcal{H}_{p}$ defined in Eq.~\eqref{eq:pol_set} excluding $\ell$, $\mathcal{A}_{m}\geq0$ and $\varphi_{m}\in[0,2\pi)$. As an example, suppose we choose the plus mode as the basis mode in the scalar-tensor hypothesis, the parallel beam pattern vector $\boldsymbol{f}_{\parallel}$ is then
\begin{equation}
	\boldsymbol{f}_{\parallel} = \boldsymbol{f}_{+} + \sum_{m=\{\times,b,l\}}\mathcal{A}_{m}e^{i\varphi_{m}}\boldsymbol{f}_{m}\,.
\end{equation}
The beam pattern matrix $\boldsymbol{F}$ is
\begin{equation}
\begin{split}
	&\boldsymbol{F}(\alpha,\delta,\psi,\boldsymbol{\mathcal{A}},\boldsymbol{\varphi},t_{\text{event}}) \\
	&= 
	\begin{bmatrix}
		\boldsymbol{f}_{\parallel}
	\end{bmatrix} \\
	&=
	\begin{bmatrix}
		\boldsymbol{f}_{+} + \sum\limits_{m=\{\times,b,l\}}\mathcal{A}_{m}e^{i\varphi_{m}}\boldsymbol{f}_{m}
	\end{bmatrix}
\end{split}
\end{equation}
where $\boldsymbol{\mathcal{A}}=\{A_{\times}, A_{b}, A_{l}\}$ and $\boldsymbol{\varphi} = \{\varphi_{\times},\varphi_{b},\varphi_{l}\}$. The null operator is then constructed with Eq.~\eqref{eq:P_def}. One should notice that now it is also a function of $\boldsymbol{\mathcal{A}}$ and $\boldsymbol{\varphi}$ in addition to $\alpha$, $\delta$, $\psi$ and $t_{\text{event}}$. The polarization angle $\psi$ is now relevant in the construction of $\boldsymbol{P}$ since we can no longer factor out $\psi$ from $\boldsymbol{F}$ in constrast to the generic null projection discussed in Appendix~\ref{app:pol_angle}. One exception is the scalar hypothesis which involves $\boldsymbol{f}_{b}$ and $\boldsymbol{f}_{l}$, but they are linearly dependent so we could choose either one of them. With a single beam pattern vector, the $\psi$ can be factored out and the construction of null operator is therefore independent of $\psi$. Other polarization hypotheses are formulated with the same manner by including the relevant $\boldsymbol{f}_{m}$ into $\boldsymbol{f}_{\parallel}$.

\subsubsection{Two-basis-mode ($L=2$) analysis}

In the $L=2$ analysis, we choose two polarization modes as the basis. The beam pattern matrix $\boldsymbol{F}$ of the polarization hypothesis $\mathcal{H}_{p}$ with the basis modes $k$ and $\ell$ is defined by
\begin{equation}
\label{eq:L2_pol}
	\begin{split}
		&\boldsymbol{F}(\alpha, \delta, \psi, \boldsymbol{\mathcal{A}}, \boldsymbol{\varphi}, t_{\text{event}}) \\
		&=\begin{bmatrix}
			\boldsymbol{f}_{\parallel,k} & \boldsymbol{f}_{\parallel,\ell}
		\end{bmatrix}
	\end{split}
\end{equation}
where
\begin{equation}
	\boldsymbol{f}_{\parallel,k}=\boldsymbol{f}_{k} + \sum\limits_{m\in\mathbb{F}_{p} \setminus \{k,\ell\} }
	\mathcal{A}_{1,m}e^{i\varphi_{1,m}}\boldsymbol{f}_{m}
\end{equation}
and
\begin{equation}
	\boldsymbol{f}_{\parallel,\ell} = \boldsymbol{f}_{\ell} + \sum\limits_{m\in\mathbb{F}_{p} \setminus \{k,\ell\} }
	\mathcal{A}_{2,m}e^{i\varphi_{2,m}}\boldsymbol{f}_{m}
\end{equation}
where $\mathbb{F}_{p} \setminus \{k,\ell\}$ is the set of labels of polarizations in hypothesis $\mathcal{H}_{p}$ defined in Eq.~\eqref{eq:pol_set} excluding $k$ and $\ell$, $\mathcal{A}_{\{1,2\},m}\geq0$ and $\varphi_{\{1,2\},m}\in[0,2\pi)$. In principle we could choose any two polarization modes as the basis, but we shall see from the non-GR waveforms of e.g.\ Brans-Dicke theory, Rosen's theory and Lightman-Lee theory presented in Ref.~\cite{Chatziioannou_2012}, the dipole radiation only enters the non-tensorial polarizations. Therefore, the more sensible choice is to choose one tensorial mode and one non-tensorial mode as the basis. As a comparison with the example in the $L=1$ analysis, we shall again take the scalar-tensor hypothesis as an example. Suppose we choose the plus mode and the scalar breathing mode as the basis, the two parallel beam pattern vectors are
\begin{equation}
	\boldsymbol{f}_{\parallel,+} = \boldsymbol{f}_{+} + \sum_{m=\{\times,l\}}\mathcal{A}_{1,m}e^{i\varphi_{1,m}}\boldsymbol{f}_{m}
\end{equation}
and
\begin{equation}
	\boldsymbol{f}_{\parallel,b} = \boldsymbol{f}_{b} + \sum_{m=\{\times,l\}}\mathcal{A}_{2,m}e^{i\varphi_{2,m}}\boldsymbol{f}_{m}\,.
\end{equation}
The beam pattern matrix is therefore
\begin{equation}
\begin{split}
	\boldsymbol{F}(\alpha,\delta,\psi,\boldsymbol{\mathcal{A}},\boldsymbol{\varphi},t_{\text{event}})
	=
	\begin{bmatrix}
		\boldsymbol{f}_{\parallel,+} & \boldsymbol{f}_{\parallel,b}
	\end{bmatrix}
\end{split}
\end{equation}
where $\boldsymbol{\mathcal{A}}=\{A_{1,\times}, A_{1,l}, A_{2,\times}, A_{2,l}\}$ and $\boldsymbol{\varphi} = \{\varphi_{1,\times},\varphi_{1,l}, \varphi_{2,\times},\varphi_{2,l}\}$. The null operator is then constructed with Eq.~\eqref{eq:P_def}. The scalar hypothesis is not defined in the $L=2$ analysis since $\boldsymbol{f}_{b}$ and $\boldsymbol{f}_{l}$ are linearly dependent. For the tensor hypothesis and the vector hypothesis, since they only have two polarization modes, they are fully characterized by themselves without requiring the parameters $\boldsymbol{\mathcal{A}}$ and $\boldsymbol{\varphi}$, and therefore $\psi$ is irrelevant in the construction of the null operator as discussed in Appendix~\ref{app:pol_angle}.

The extension to even more basis modes ($L>2$) is trivial. Since we are most interested in detecting GR violation, but the tensor hypothesis is not defined in $L>2$ analyses, so the $L=1$ and the $L=2$ analyses are sufficient in terms of probing deviation from GR.

\subsection{Bayesian model selection}

Since the model parameters $\boldsymbol{\theta}=\{\alpha$, $\delta, \psi,\boldsymbol{\mathcal{A}},\boldsymbol{\varphi}\}$ are not known (we take the event time $t_{\text{event}}$ from the search pipelines as known), we adopt a Bayesian analysis to marginalize the unknown parameters to obtain the Bayesian evidence of the competing hypotheses.

To improve the sensitivity, we analyze the data in the time-frequency domain and target the region spanned by the candidate signal through time-frequency clustering. We use the Wilson-Daubechies-Meyer (WDM) time-frequency transform \cite{Necula_2012} because of its superior spectral leakage control. Given a multi-detector discrete time series $\boldsymbol{x}\in\mathbb{R}^{D\times N}$ where $D$ is the number of detectors and $N$ is the
number of time bins, denote the time-frequency transform as $\textbf{TF}:\mathbb{R}^{D\times N}\rightarrow\mathbb{R}^{D\times J\times K}$ where $J$ is the number of time bins and $K$ is the number of frequency bins which performs the WDM time-frequency transform on the time series of each detector independently. The corresponding time-frequency representation is denoted with a subscript as $\boldsymbol{x}_{\text{TF}}$.

One should notice that the construction of the null operator involves the noise-weighed beam pattern function matrix $\boldsymbol{F}_{w}$ in Eq.~\eqref{eq:Fw}. The noise PSD of LIGO-Virgo interferometers typically exhibits sharp spectral lines due to the resonance of wires in suspension, electrical supply power line, and injected calibration lines \cite{Littenberg_2015}. To reduce the spectral leakage, we compute the residual $\tilde{\boldsymbol{z}}$ in Eq.~\eqref{eq:residual_start} in the frequency domain which has a higher frequency resolution than the time-frequency representation. Then, we perform the inverse Fourier transform $\mathcal{F}^{-1}$ to transform the residual back to the time domain and then transform the time domain residual to the time-frequency domain by $\textbf{TF}$. As shown in Eq.~\eqref{eq:residual_start}, when the null operator $\boldsymbol{P}$ is constructed with the correct parameters, the residual $\tilde{\boldsymbol{z}}$ is only noise in the lower dimensional space regardless of $\tilde{\boldsymbol{h}}$. The time-frequency representation of the residual $\boldsymbol{z}_{\text{TF}}$ is also only noise with the null energy $E_{\text{null}}$ \cite{Sutton_2010} defined as
\begin{equation}
	\label{eq:null_energy_tf}
    E_{\text{null}}(\boldsymbol{z}_{\text{TF}}) = \sum_{j=1}^{D}\sum_{\{k,l\}\in\mathcal{C}}\norm{z_{\text{TF},j}[k,l]}^{2}
\end{equation}
where $z_{\text{TF},j}[k,l]$ denotes the coefficient of the time-frequency representation $\boldsymbol{z}_{\text{TF}}$ corresponding to the $k^{\text{th}}$ time index and $l^{\text{th}}$ frequency index of the residual of the $j^{\text{th}}$ detector, and $\mathcal{C}$ denotes the set of time-frequency indices that the candidate signal occupies on the time-frequency plane follows the $\chi^{2}$ distribution with a degree of freedom $(D-L)N$ where $L$ is the number of basis modes and $N$ is the number of time-frequency pixels that the candidate signal spans when the null operator is correctly constructed. The degree of freedom is half of that in Ref.~\cite{Sutton_2010} since the WDM coefficients are real but in Ref.~\cite{Sutton_2010} Fourier transform is used and the real and imaginary components double the degree of freedom.

The likelihood is then given by
\begin{equation}
    p(\boldsymbol{d}|\alpha,\delta,\psi,\boldsymbol{\mathcal{A}},\boldsymbol{\varphi},\mathcal{H}) = \chi^{2}_{\text{DoF}}(E_{\text{null}}(\boldsymbol{z}_{\text{TF}}))
\end{equation}
where $\boldsymbol{z}_{\text{TF}}=\textbf{TF}(\mathcal{F}^{-1}(\boldsymbol{P}_{\mathcal{H}}(\tilde{\boldsymbol{d}};\alpha,\delta,\psi,\boldsymbol{\mathcal{A}},\boldsymbol{\varphi})))$, $\boldsymbol{P}_{\mathcal{H}}$ denotes the null operator constructed with the $\boldsymbol{f}_{m}$ implied by the hypothesis $\mathcal{H}$ and the number of basis modes $L$, and $\chi_{\text{DoF}}^{2}(\cdot)$ is the $\chi^{2}$ probability density function with the degree of freedom $\text{DoF}$ implied by the hypothesis $\mathcal{H}$. The Bayesian evidence is then
evaluated by marginalizing all parameters
\begin{equation}
\begin{split}
    &p(\boldsymbol{d}|\mathcal{H}) \\
    &= \int p(\boldsymbol{d}|\alpha,\delta,\psi,\boldsymbol{\mathcal{A}},\boldsymbol{\varphi},\mathcal{H})p(\alpha,\delta,\psi,\boldsymbol{\mathcal{A}},\boldsymbol{\varphi}|\mathcal{H})d\alpha d\delta d\psi d\boldsymbol{\mathcal{A}}d\boldsymbol{\varphi}
\end{split}
\label{eq:evidence}
\end{equation}
where $p(\alpha,\delta,\psi,\boldsymbol{A},\boldsymbol{\varphi}|\mathcal{H})$ is the prior distribution of the parameters given $\mathcal{H}$. The posterior odds between $\mathcal{H}_{1}$ and $\mathcal{H}_{2}$ defined as
\begin{equation}
\begin{split}
    \mathcal{O}_{\mathcal{H}_{2}}^{\mathcal{H}_{1}} &= \frac{p(\mathcal{H}_{1}|\boldsymbol{d})}{p(\mathcal{H}_{2}|\boldsymbol{d})}\\
    &=\frac{p(\boldsymbol{d}|\mathcal{H}_{1})}{p(\boldsymbol{d}|\mathcal{H}_{2})}\times\frac{p(\mathcal{H}_{1})}{p(\mathcal{H}_{2})}
\end{split}
\end{equation}
that decribes the ratio of probabilities between the two hypotheses $\mathcal{H}_{1}$ and $\mathcal{H}_{2}$ given the observed data $\boldsymbol{d}$, $\frac{p(\boldsymbol{d}|\mathcal{H}_{1})}{p(\boldsymbol{d}|\mathcal{H}_{2})}$ is the ratio of model evidences and is also called the Bayes factor, and $\frac{p(\mathcal{H}_{1})}{p(\mathcal{H}_{2})}$ is the prior odds between the hypotheses
$\mathcal{H}_{1}$ and $\mathcal{H}_{2}$ which describes the a priori belief of the ratio of probabilities of the two hypotheses. When we are ignorant about the relative probability between the competing hypothesis, we take the prior odds to be $1$, and then the posterior odds would be equal to the Bayes factor. In the following sections, we use the Bayes factor as the detection statistic. It should be understood that Eq.~\eqref{eq:evidence} is presented for generality,
$\{\psi,\boldsymbol{A},\boldsymbol{\varphi}\}$ are not present in the integral when the hypothesis $\mathcal{H}$ does not have the extra parameterization.

As a follow-up analysis, in the analysis pipeline, we require the time-frequency cluster found to span across the event time reported by the search pipeline. This could fail when the signal is too weak, and there are loud noise transients around the event. If the cluster does not span across the event time, the analysis is performed in the frequency domain, and the whole data chunk is being analyzed. The null energy is then computed with the frequency domain residual $\tilde{\boldsymbol{z}}$
\begin{equation}
	E_{\text{null}}(\tilde{\boldsymbol{z}}) = \sum_{j=1}^{D}\sum_{k=1}^{K}\norm{\tilde{z}_{j}[k]}^{2}
\end{equation}
where $\tilde{z}_{j}[k]$ denotes the residual at frequency bin $k$ of the $j^{\text{th}}$ detector, $D$ denotes the number of detectors and $K$ denotes the number of frequency bins. $2E_{\text{null}}$ then follows the $\chi^{2}$ distribution with degree of freedom $2(D-M)K$ when the null operator is correctly constructed.

\subsection{The orthogonal component $\tilde{h}_{\perp}$}
\label{sec:orthogonal}

One would still need to pay attention to the orthogonal polarization component $\tilde{h}_{\perp}$ with the partial null projection. In terms of testing GR, the orthogonal polarization component must not be significant in the tensor hypothesis, and otherwise, this could lead to false GR violation. For CBC signals, the plus- and the cross-polarization modes in the dominant 22-mode only differ by the amplitude (due to the inclination angle) and the phase by $\pi/2$ which are frequency-independent \cite{creighton_anderson_2011}. The addition of the higher-order harmonics only adds subdominant corrections. The plus mode and the cross mode are still well described by a single basis mode, and the orthogonal component is vanishingly small. In the example of the $L=1$ analysis presented in Sec.~\ref{sec:L_1}, the orthogonal component would be significant if the dipole radiation only enters the non-tensorial polarizations but not the tensorial polarizations like the Brans-Dicke theory \cite{Chatziioannou_2012}. We would argue that detecting the non-tensorial component parallel to the basis modes is sufficient for us to detect a GR violation. However, we also want to know how much uncaptured $\tilde{h}_{\perp}$ are in the data to understand how well the model can explain the data. This is important for us to understand the systematics of the ranking of the non-tensor hypotheses if we observe a deviation from GR.

With the same spirit of the residuals test \cite{PhysRevD.100.104036,theligoscientificcollaboration2020tests}, we analyze whether the null energy defined in 	Eq.~\eqref{eq:null_energy_tf} with the maximum-likelihood parameters are consistent with noise or not. The consistency could be quantified using the \textit{plug-in p-value} \cite{Demortier:2007zz} defined by
\begin{equation}
	p_{\text{plug-in}}(E_{\text{null}}) = \int_{E_{\text{null}}}^{\infty}\chi^{2}_{\text{DoF}}(E)dE
\end{equation}
where $E_{\text{null}}$ is the null energy with the maximum-likelihood parameters. One should notice that the plug-in p-value does not distribute uniformly between $0$ and $1$. Instead, it should always be around $p_{\text{plug-in}}(\text{DoF}-2)$ (assume $\text{DoF}\geq2$) since the mode of a $\chi^{2}$ distribution with a degree of freedom $\text{DoF}$ is $\text{DoF}-2$ that is the $E_{\text{null}}$ with the highest possible likelihood. A significantly smaller $p_{\text{plug-in}}(E_{\text{null}})$ than the $p_{\text{plug-in}}(\text{DoF}-2)$ indicates the presence of an orthogonal polarization component. The $p_{\text{plug-in}}$ could therefore be served as a diagnostic tool, but one shoule be reminded that it cannot be interpreted as the usual p-value since it does not distriute uniformly between $0$ and $1$ under the null hypothesis. A more interpretable statistic could be obtained using the double bootstrap method to calculate the adjusted plug-in p-value \cite{Demortier:2007zz,davison_hinkley_1997} that distributes uniformly between $0$ and $1$.

\subsection{Calibration errors}

Calibration errors are present in instruments \cite{Cahillane_2017,Accadia_2010}. Although a previous study \cite{Vitale_2012} shows that calibration-induced errors of Advanced LIGO and Advanced Virgo are not a significant detriment to accurate parameter estimation, including the effects of calibration errors has been a standard practice in the LIGO Scientific Collaboration and the Virgo Collaboration \cite{PhysRevX.9.031040,abbott2020gwtc2} to improve the accuracy of results and for completeness. The details of including effects of calibration errors into the framework are discussed in Appendix~\ref{app:cal_error}.

\section{Results}
\label{sec:results}
We experiment on tensor, vector, scalar, tensor-scalar, tensor-vector, vector-scalar and tensor-vector-scalar injections to validate the method. In this section, we start with the ad hoc non-tensorial injections which non-tensorial components are generated by projecting $\tilde{h}_{+}(f)$ and $\tilde{h}_{\times}(f)$ onto the non-tensorial beam pattern functions. The polarization modes can then be described well with a single basis mode. This serves as testing the methods when there is no orthogonal polarization component. We then experiment on the more realistic non-GR waveforms with both cases when the orthogonal polarization component is present and absent respectively.

\subsection{Nested sampling configurations}
The Bayesian model evidences are computed with nested sampling using \texttt{MultiNest} \cite{Feroz_2008,Feroz_2009,Feroz_2019}. The prior of source sky position $\hat{\Omega} = (\alpha$, $\delta)$ is taken to be uniform over the sky sphere. The prior of polarization angle $\psi$ is taken to be uniform over $[0,\pi]$. The prior of each relative amplitude in $\boldsymbol{\mathcal{A}}$ is taken to be uniform over $[0,2]$. The prior of each relative phase in $\boldsymbol{\varphi}$ is taken to be uniform over $[0,2\pi]$. 1024 live points are used. The sampling efficiency is set to be $0.3$ as recommended in the GitHub repository of \texttt{MultiNest} \cite{multinest_github_2019}.

\subsection{Ad hoc injections}
\label{sec:ad_hoc_injections}
\subsubsection{Mock data preparation}
\label{subsec:mock_data}
One should expect to observe a stronger model preference for the non-tensor hypotheses when the underlying non-tensorial signal has a higher signal-to-noise ratio (SNR). Also, one should expect to observe a stronger model preference for the tensor hypothesis when the underlying tensorial signal has a higher SNR. Therefore, we generate tensorial and non-tensorial injections, and check whether the correct hypothesis is more favored. We fix the noise realization and every waveform parameter except the luminosity distance for scaling to the targeted SNR. We use the \texttt{IMRPhenomPv2} \cite{PhysRevD.93.044006,PhysRevD.93.044007,PhysRevLett.113.151101} waveform model to generate the binary black hole (BBH) waveforms $h_{+}(t)$ and $h_{\times}(t)$ with a sampling rate $f_{s} = 2048\text{ Hz}$ and a frequency lower cut $f_{\text{low}}=20\text{ Hz}$. The component masses are set to be $m_{1}=m_{2}=20M_{\odot}$. The spins,  inclination angle, coalescence phase and polarization angle are set to be $0$. The geocentric GPS time is set to be $1282107824$ and the right ascension and declination of the source location are set to be $\alpha=2.72$ and $\delta=-0.36$ respectively. We generate the non-tensorial signals by projecting $h_{+}(t)$ and $h_{\times}(t)$ onto the non-tensorial beam pattern functions. The waveforms are projected onto the Hanford-Livingston-Virgo (HLV) detector network, and injected into simulated Gaussian noise with the advanced LIGO and advanced Virgo design sensitivities, or more specifically \texttt{aLIGODesignSensitivityP1200087} and \texttt{AdVDesignSensitivityP1200087} \cite{Abbott_2020} respectively.

We first present the results of the analysis on the injections with pure polarizations i.e.\ pure tensorial, pure vectorial, and pure scalar signal. The injected signals are scaled to 20 different network SNRs equally spaced between 10 and 100. The single detector SNR is defined by
\begin{equation}
	\rho = \sqrt{4\int_{0}^{\infty} \frac{\norm{\tilde{s}(f)}^{2}}{S(f)}df}
\end{equation}
where $\tilde{s}(f)$ is the Fourier transform of the observed signal and $S(f)$ is the one-sided PSD of detector noise, and the $N$-detector network SNR is defined by
\begin{equation}
	\rho_{\text{net}} = \sqrt{\sum_{j=1}^{N}\rho_{j}^{2}}\,.
\end{equation}
We then present the results of the analysis on the injections with mixed polarizations. We study the case when the signal is dominantly tensorial with a non-tensorial correction. We fix the network SNR of the injections to be $100$, and vary the strength of the non-tensorial component. Lastly, we also present the results of the analysis on vector-scalar injections for completeness.

\subsubsection{Pure polarizations}
The scalar signal is generated by
\begin{equation}
	s_{S}(t) = F_{b}h_{+}(t) + F_{l}h_{\times}(t)
\end{equation}
and a vector signal is generated by
\begin{equation}
	s_{V}(t) = F_{x}h_{+}(t) + F_{y}h_{\times}(t)\,.
\end{equation}

\begin{figure}
	\includegraphics[width=\linewidth]{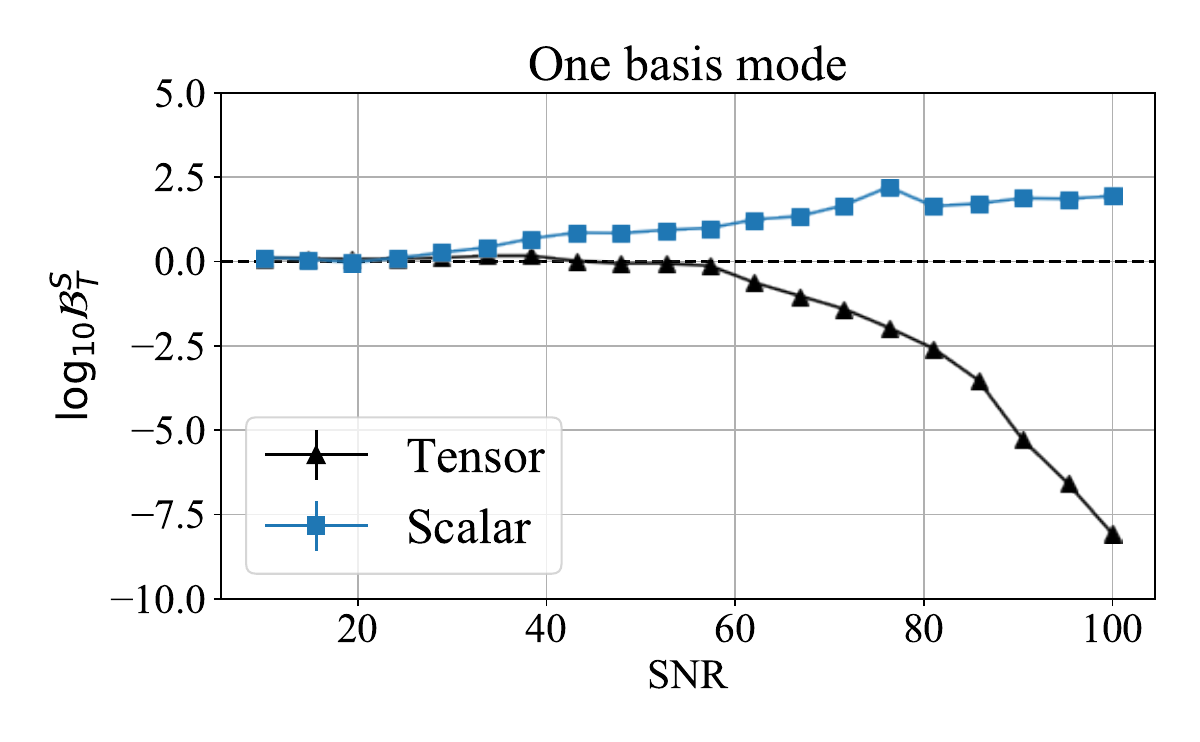}
	\caption{The plot shows the variation of $\log_{10}\mathcal{B}_{T}^{S}$ with varying SNR. The signals are injected into the Hanford-Livingston-Virgo 3-detector network. The squares denote the results of analysis of scalar injections, and the triangles denote the results of analysis of tensor injections. The error bars are the standard error of $\log_{10}\mathcal{B}_{T}^{S}$ that due to the error of evidence estimation of the sampler. The error bars are too small to be seen. Since the scalar breathing and scalar longitudinal beam pattern functions are linearly dependent, the scalar hypothesis is only defined with one basis mode. The tensor hypothesis here takes the $+$ mode as the basis. The beam pattern matrix of the hypotheses is defined in Eq.~\eqref{eq:L1_pol}. The true polarization hypotheses are more favored with an increasing SNR.}
	\label{fig:result_S_T}
\end{figure}

Fig.~\ref{fig:result_S_T} shows the results of $\log_{10}$ Bayes factor of scalar hypothesis $\mathcal{H}_{S}$ against tensor hypothesis $\mathcal{H}_{T}$ ($\log_{10}\mathcal{B}_{T}^{S}$) with scalar injections and tensor injections. Since the scalar breathing beam pattern function and scalar longitudinal beam pattern function are linearly dependent, the scalar hypothesis is only defined with one basis mode. Consequently, the competing hypotheses also have to be constructed with one basis mode. The competing tensor hypothesis is hence defined with the $+$ mode chosen to be the basis.

\begin{figure}
	\includegraphics[width=\linewidth]{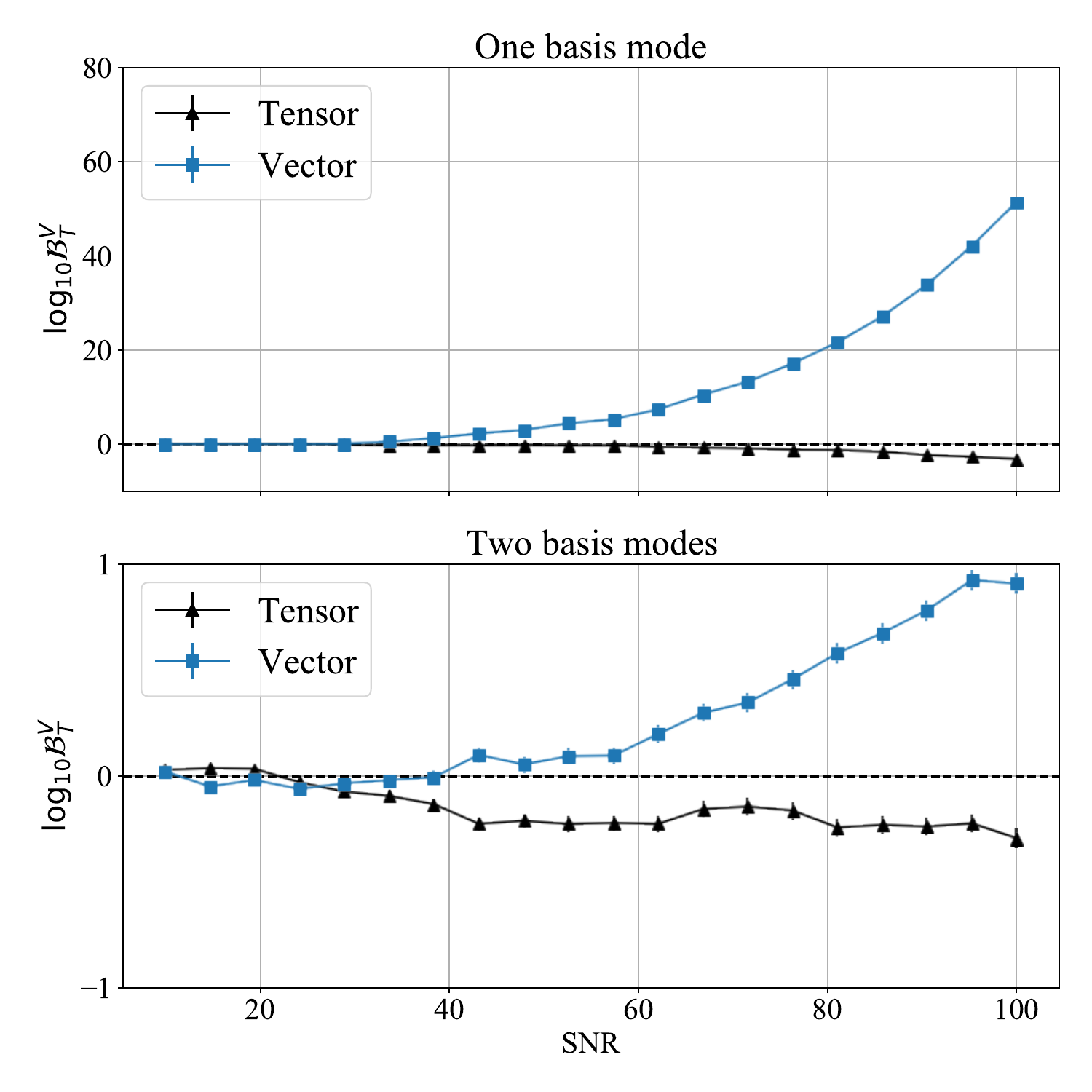}
	\caption{The plots show the variation of $\log_{10}\mathcal{B}_{T}^{V}$ with varying SNR. The signals are injected into the Hanford-Livingston-Virgo 3-detector network. The squares denote the results of analysis of vector injections, and the triangles denote the results of analysis of tensor injections. The error bars are the standard error of $\log_{10}\mathcal{B}_{T}^{V}$ that due to the error of evidence estimation of the sampler. The error bars are too small to be seen. The upper panel shows the results with the $x$ mode as the basis component for the vector hypothesis, and the $+$ mode as the basis component for the tensor hypothesis. The beam pattern matrix of the one-basis-mode analysis is defined in Eq.~\eqref{eq:L1_pol}. The lower panel shows the results with the $x$ and the $y$ modes as the basis components for the vector hypothesis, and the $+$ mode and the $\times$ modes as the basis components for the tensor hypothesis. The beam pattern matrix of the two-basis-mode analysis is defined in Eq.~\eqref{eq:L2_pol}. The true polarization hypotheses are more favored with an increasing SNR.}
	\label{fig:result_V_T}
\end{figure}

Fig.~\ref{fig:result_V_T} shows the results of $\log_{10}$ Bayes factor of vector hypothesis $\mathcal{H}_{V}$ against tensor hypothesis $\mathcal{H}_{T}$ ($\log_{10}\mathcal{B}_{T}^{S}$) with vector injections and tensor injections. The upper panel shows the results with one basis mode. $x$ mode is chosen as the basis for the vector hypothesis and $+$ mode is chosen as the basis for the tensor hypothesis. The lower panel shows the results with two basis modes. Since both vector polarizations and tensor polarizations have two polarization modes, $x$ and $y$ modes are the only choices of basis for vector hypothesis, and $+$ and $\times$ modes are the only choices of basis for the tensor hypothesis.

The error bars denote the $\pm1$ sigma error of $\log_{10}\mathcal{B}$ estimated by $\sqrt{\sigma_{1}^{2}+\sigma_{2}^{2}}$ where $\sigma_{1,2}$ are the one sigma error of the competing model log evidences from the \texttt{MultiNest} \cite{Feroz_2008,Feroz_2009,Feroz_2019} outputs after converting from base $e$ to base $10$. The plots show a tendency to be in more favor of the true polarization hypotheses with an increasing SNR.

\subsubsection{Tensor polarizations with non-tensorial corrections}
We study the case when the polarization content is dominantly tensorial with non-tensorial corrections. We should expect to observe a stronger model preference for the non-tensor hypotheses with a stronger non-tensorial component. We perform the analysis on tensor-scalar, tensor-vector and tensor-vector-scalar injections with different strengths of non-tensorial components. A tensor-scalar signal is generated by
\begin{equation}
	s_{TS}(t) = F_{+}h_{+}(t) + F_{\times}h_{\times}(t) + \mathcal{A}(F_{b}h_{+}(t) + F_{l}h_{\times}(t))
\end{equation}
where $\mathcal{A}$ denotes the relative amplitude of the scalar modes to the tensor modes. A tensor-vector signal is generated by
\begin{equation}
	s_{TV}(t) = F_{+}h_{+}(t) + F_{\times}h_{\times}(t) + \mathcal{A}(F_{x}h_{+}(t) + F_{y}h_{\times}(t))
\end{equation}
where $\mathcal{A}$ denotes the relative amplitude of the vector modes to the tensor modes. A tensor-vector-scalar signal is generated by
\begin{equation}
	\begin{split}
	s_{TVS}(t) &= F_{+}h_{+}(t) + F_{\times}h_{\times}(t) \\
	&+ \mathcal{A}(F_{x}h_{+}(t) + F_{y}h_{\times}(t) + F_{b}h_{+}(t) + F_{l}h_{\times}(t))
	\end{split}
\end{equation}
where $\mathcal{A}$ denotes the relative amplitude of the non-tensorial modes to the tensor modes. Signals with $\mathcal{A}=\{0,0.2,0.4,0.6,0.8,1.0\}$ are generated and injected into the HLV detector network with the network SNR fixed to be $100$.

\begin{figure}
	\includegraphics[width=\linewidth]{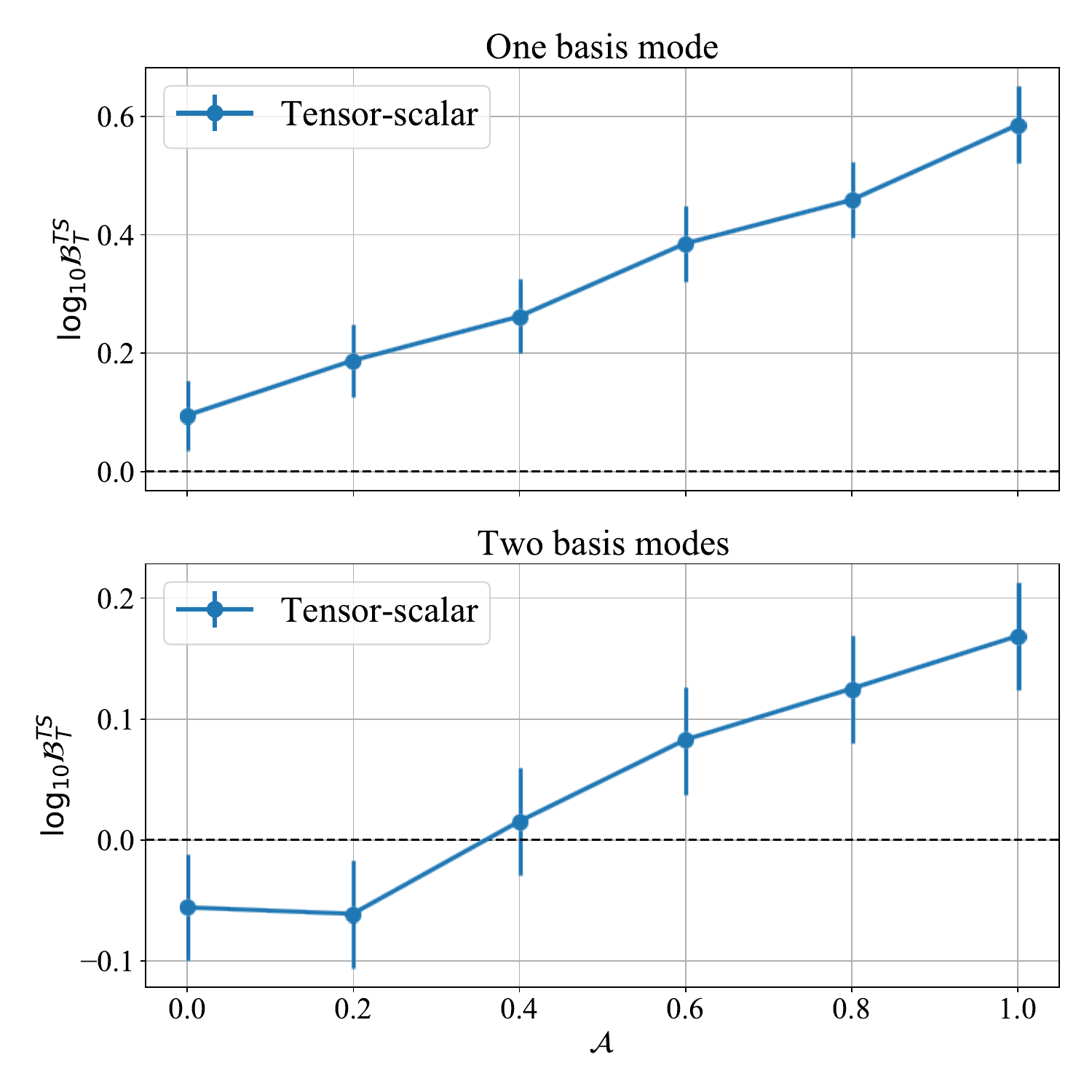}
	\caption{The plots show the variation of $\log_{10}\mathcal{B}_{T}^{TS}$ with a tensor-scalar injection of a varying relative strength $\mathcal{A}$ of the scalar polarization mode relative to the tensorial polarization mode. The signals are injected into the Hanford-Livingston-Virgo 3-detector network and the network SNR is fixed to be $100$ for all injections. The error bars are the standard error of $\log_{10}\mathcal{B}_{T}^{TS}$ that due to the error of evidence estimation of the sampler. The upper panel shows the results using the $+$ mode as the basis component for both of the tensor hypothesis $\mathcal{H}_{T}$ and the tensor-scalar hypothesis $\mathcal{H}_{TS}$. The lower panel shows the results using the $+$ mode and the $\times$ mode as the basis components for both hypotheses. The tensor-scalar hypothesis is more favored when the strength of the scalar component is stronger.}
	\label{fig:result_TS_T}
\end{figure}

\begin{figure}
	\includegraphics[width=\linewidth]{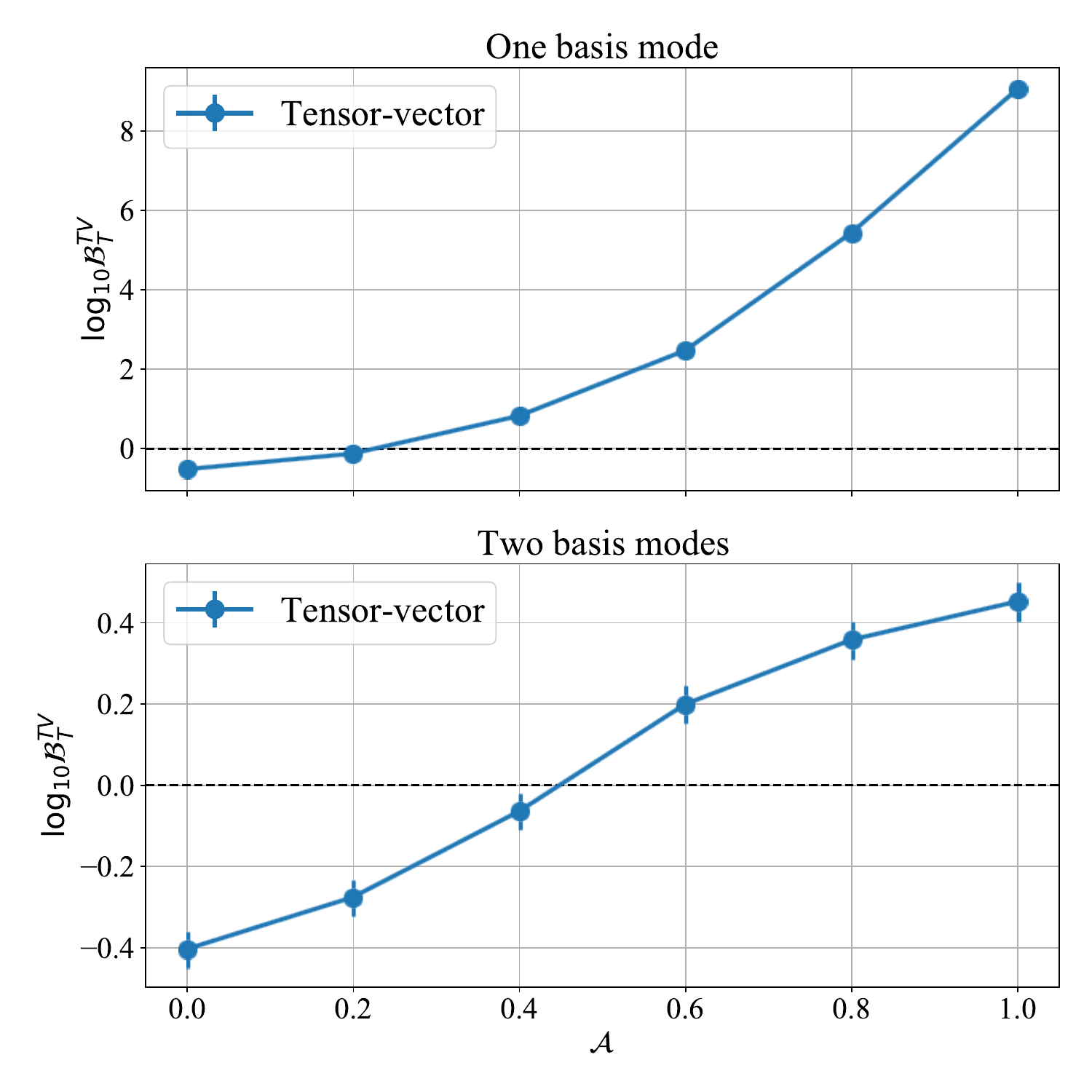}
	\caption{The plots show the variation of $\log_{10}\mathcal{B}_{T}^{TV}$ with a tensor-vector injection of a varying relative strength $\mathcal{A}$ of the vector polarization mode relative to the tensorial polarization mode. The signals are injected into the Hanford-Livingston-Virgo 3-detector network and the network SNR is fixed to be $100$ for all injections. The error bars are the standard error of $\log_{10}\mathcal{B}_{T}^{TV}$ that due to the error of evidence estimation of the sampler. The error bars on the upper panel are too small to be seen. The upper panel shows the results using the $+$ mode as the basis component for both of the tensor hypothesis $\mathcal{H}_{T}$ and the tensor-vector hypothesis $\mathcal{H}_{TV}$. The lower panel shows the results using the $+$ mode and the $\times$ mode as the basis components for both hypotheses. The tensor-vector hypothesis is more favored when the strength of the vector component is stronger.}
	\label{fig:result_TV_T}
\end{figure}

\begin{figure}
	\includegraphics[width=\linewidth]{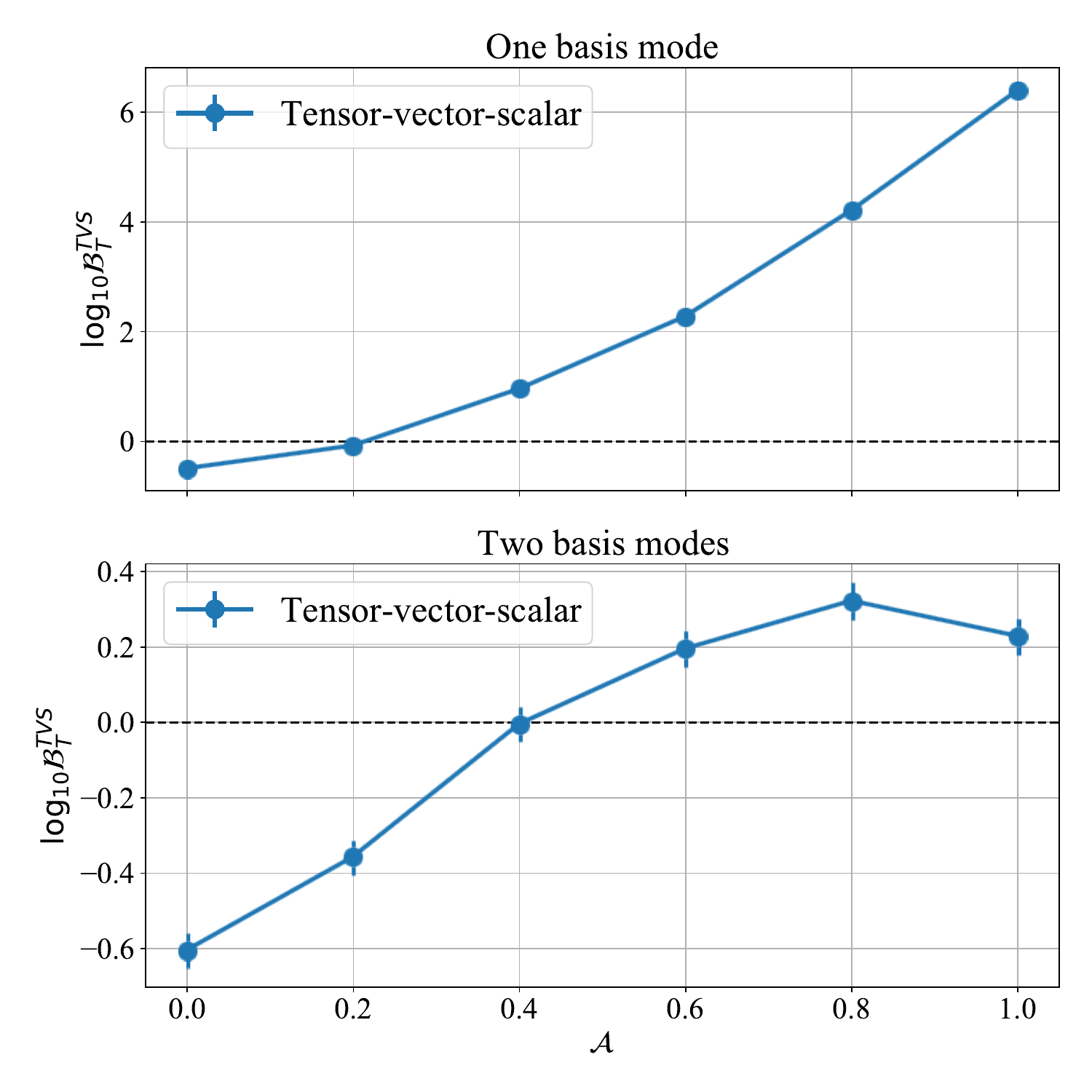}
	\caption{The plots show the variation of $\log_{10}\mathcal{B}_{T}^{TVS}$ with a tensor-vector-scalar injection of a varying relative strength $\mathcal{A}$ of the vector and scalar polarization modes relative to the tensorial polarization mode. The signals are injected into the Hanford-Livingston-Virgo 3-detector network and the network SNR is fixed to be $100$ for all injections. The error bars are the standard error of $\log_{10}\mathcal{B}_{T}^{TVS}$ that due to the error of evidence estimation of the sampler. The error bars on the upper panel are too small to be seen. The upper panel shows the results using the $+$ mode as the basis component for both of the tensor hypothesis $\mathcal{H}_{T}$ and the tensor-vector-scalar hypothesis $\mathcal{H}_{TVS}$. The lower panel shows the results using the $+$ mode and the $\times$ mode as the basis components for both hypotheses. The tensor-vector-scalar hypothesis is more favored when the strength of the non-tensorial component is stronger.}
	\label{fig:result_TVS_T}
\end{figure}

The results with the tensor-scalar injections, the tensor-vector injections and the tensor-vector-scalar injections are shown in Fig.~\ref{fig:result_TS_T}, Fig.~\ref{fig:result_TV_T} and Fig.~\ref{fig:result_TVS_T} respectively. For comparison, we also perform the same analysis with tensor injections and the results are shown in each of the figures. The upper panels show the results of $L=1$ analysis with the $+$ mode chosen to be the basis component, and the lower panels show the results of $L=2$ analysis with the $+$ mode and the $\times$ mode chosen to be the basis components. The figures show a general trend favoring the non-tensor hypotheses with a stronger non-tensorial component. However, the tensor hypothesis is not significantly favored when the injection is pure tensorial even with a very high SNR. This is because the non-tensor hypotheses $\mathcal{H}_{TV}$ and $\mathcal{H}_{TVS}$ with $\boldsymbol{\mathcal{A}}=\boldsymbol{0}$ could also perfectly explain the data. The slight preference towards $\mathcal{H}_{T}$ is due to the penalty on the more complicated models by the Ockham's razor \cite{sivia2006data}.

\subsubsection{Vector-scalar polarizations}
Lastly, we experiment on vector-scalar injections. The vector-scalar signal is generated by
\begin{equation}
	s_{VS}(t) = F_{x}h_{+}(t)+F_{y}h_{\times}(t)+F_{b}h_{+}(t)+F_{l}h_{\times}(t)\,.
\end{equation}
\begin{figure}
	\includegraphics[width=\linewidth]{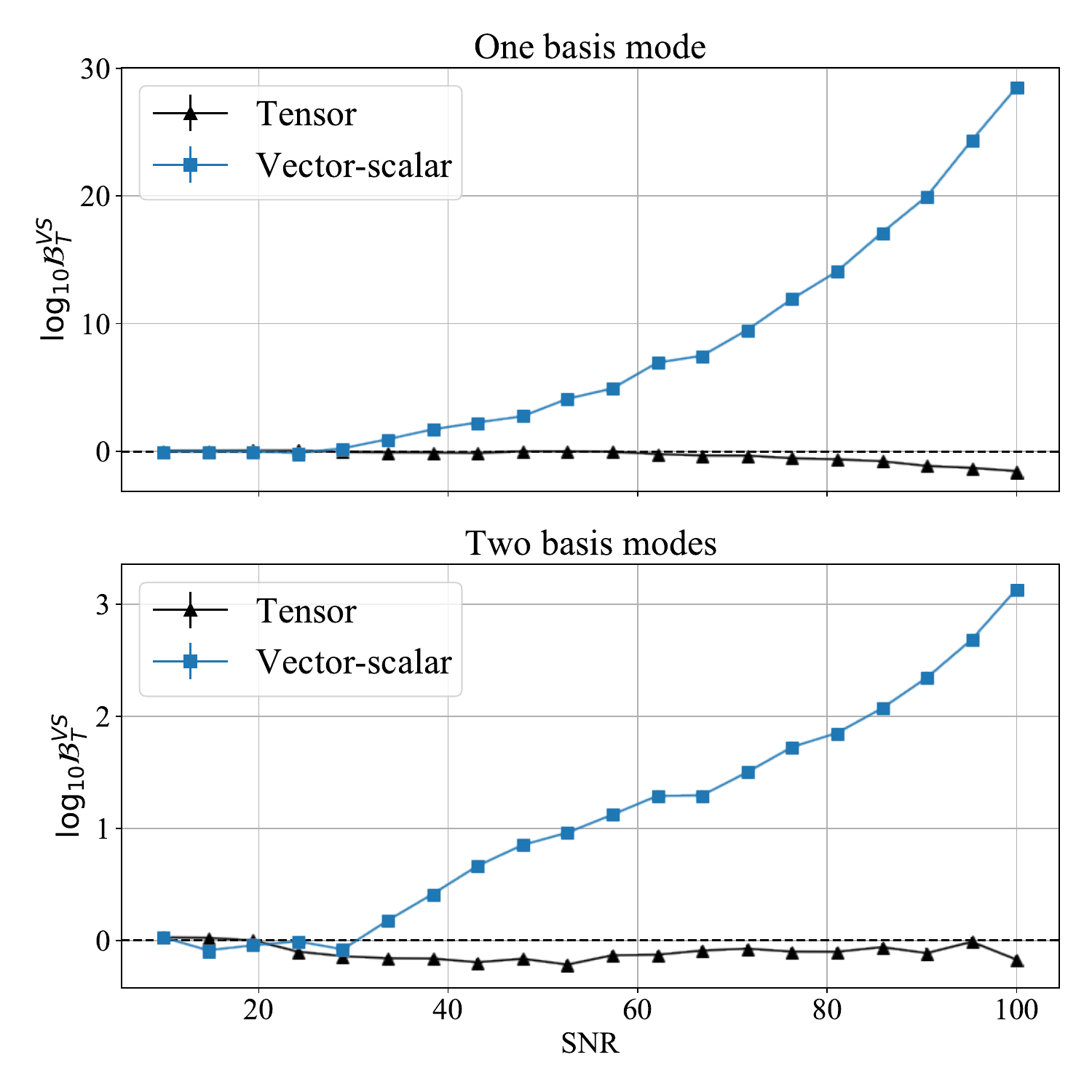}
	\caption{The plots show the variation of $\log_{10}\mathcal{B}_{T}^{VS}$ with varying SNR. The signals are injected into the Hanford-Livingston-Virgo 3-detector network. The squares denote the results of analysis of vector-scalar injections, and the triangles denote the results of analysis of tensor injections. The error bars are the standard errors of $\log\mathcal{B}_{T}^{VS}$ that due to the error of evidence estimation of the sampler. The error bars are too small to be seen in the plots. The upper panel shows the results using one basis mode with $x$ mode as the basis for $\mathcal{H}_{VS}$ and $+$ mode as the basis for $\mathcal{H}_{T}$. The beam pattern matrix of the one-basis-mode analysis is defined in Eq.~\eqref{eq:L1_pol}. The lower panel shows the results using two basis modes with $x$ and $y$ modes as the basis components for $\mathcal{H}_{VS}$ and $+$ mode and $\times$ mode as the basis components for $\mathcal{H}_{T}$. The beam pattern matrix of the two-basis-mode analysis is defined in Eq.~\eqref{eq:L2_pol}. The true polarization hypotheses are more favored with an increasing SNR.}
	\label{fig:result_VS_T}
\end{figure}
The results are shown in Fig.~\ref{fig:result_VS_T}. Similarly, the upper panel shows the results using one basis mode. $x$ mode is chosen to be the basis component for the vector-scalar hypothesis, and $+$ mode is chosen to be the basis component for the tensor hypothesis. The lower panel shows the results using two basis modes. $x$ and $y$ modes are chosen to be the basis components for the vector-scalar hypothesis, and $+$ mode and $\times$ mode are chosen to be the basis components for the tensor hypothesis. The plots show a general trend favoring the true underlying polarization models with a higher SNR.

\subsubsection{Discussion}

\begin{figure}
	\includegraphics[width=\linewidth]{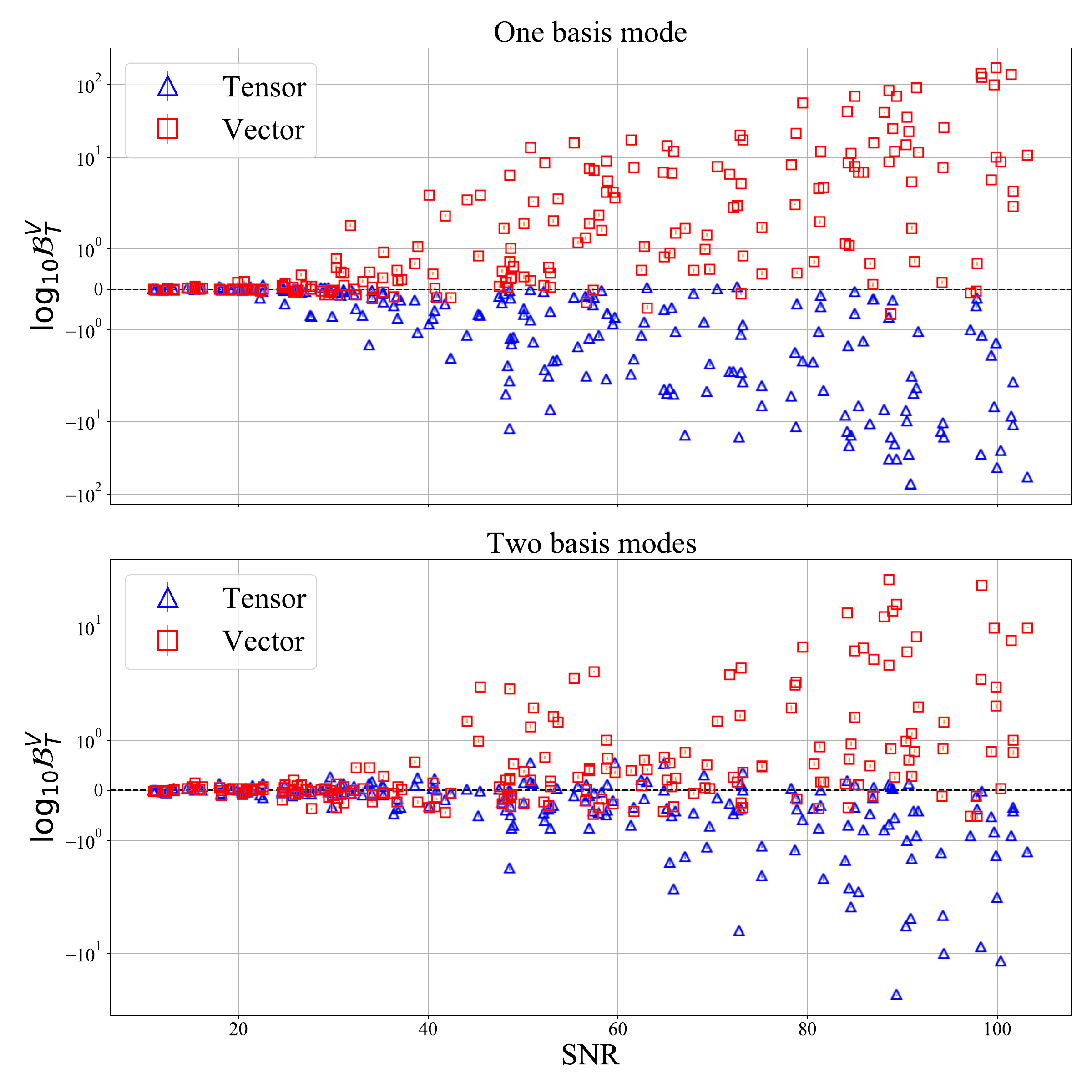}
	\caption{The plots show the distribution of $\log_{10}\mathcal{B}_{T}^{V}$ which injections are generated with randomly sampled waveform parameters and source locations. The squares denote the results of analysis of vector injections. The triangles denote the results of analysis of tensor injections. The upper panel shows the results with $+$ mode as the basis component. The beam pattern matrix of the one-basis-mode analysis is defined in Eq.~\eqref{eq:L1_pol}. The lower panel shows the results with $+$ and $\times$ modes as the basis components. The beam pattern matrix of the two-basis-mode analysis is defined in Eq.~\eqref{eq:L2_pol}.}
	\label{fig:result_V_T_band}
\end{figure}

The results suggest the formulation we propose is capable of probing all possible types of non-tensorial polarizations. One should observe that using one basis mode, in general, gives a stronger model preference for the true model than using two basis modes and this agrees with our expectation as discussed in Sec.~\ref{sec:formulation}. One should not overinterpret the figures to be stating the expected $\log_{10}\mathcal{B}$ to be observed given the relative strength of non-tensorial components and the SNR. The source location of all injections is set to a fixed position arbitrarily chosen as stated in Sec.~\ref{subsec:mock_data}, but the sensitivity to probe for polarizations significantly depends on the source location. If the source locations of the injections are uniformly sampled over the sky sphere, the trend line in the plots would instead appear as a wide band. Fig.~\ref{fig:result_V_T_band} shows an example plot of the distribution of $\log_{10}\mathcal{B}_{T}^{V}$ of vector and tensor injections in the HLV network with randomly sampled injection parameters. The component masses are sampled from a uniform distribution $\in[5,50]M_{\odot}$. The sky positions of the source are sampled from a uniform sky sphere. The inclination angles $\iota$ are sampled from a cosine distribution where $\iota\in[0,\pi]$. The coalescence phases are sampled from a uniform distribution $\in[0,2\pi]$. The polarization angles are sampled from a uniform distribution $\in[0,\pi]$. One should notice that there is a portion of high SNR injections having $\log_{10}\mathcal{B}_{T}^{V}\approx0$ due to the difficulty to distinguish between vector hypothesis and tensor hypothesis at those sky locations.

\subsection{More realistic non-GR injections}

\subsubsection{Mock data preparation}
\label{sec:non_GR_mock_data}

We experiment on the waveforms predicted by Rosen's theory \cite{PhysRevD.3.2317,ROSEN1974455} which is a bimetric theory that predicts the existence of all six polarization states. After combining the Eq.~69-71 in Ref.~\cite{Chatziioannou_2012}, the Fourier transform of the polarization modes in the stationary phase approximation is
\begin{equation}
\label{eq:rosen_waveform}
\begin{split}
	&\tilde{h}_{+}(f) = -\frac{1+\cos^{2}\iota}{2}k_{\text{R}}^{-1/3}\tilde{g}_{\text{R}}^{(2)}(f) \\
	&\tilde{h}_{\times}(f) = -i\cos\iota k_{\text{R}}^{-1/3}\tilde{g}_{\text{R}}^{(2)}(f) \\
	&\tilde{h}_{b}(f) = -\frac{\sin^{2}\iota}{2}k_{\text{R}}^{-1/3}\tilde{g}_{\text{R}}^{(2)}(f) - \frac{4}{3}\mathcal{G}\sin{\iota}k_{\text{R}}^{-1/6}\tilde{g}_{\text{R}}^{(1)}(f)\\
	&\tilde{h}_{l}(f) = -\sin^{2}\iota k_{\text{R}}^{-1/3}\tilde{g}_{\text{R}}^{(2)}(f)-\frac{4}{3}\mathcal{G}\sin\iota k_{\text{R}}^{-1/6}\tilde{g}_{\text{R}}^{(1)}(f) \\
	&\tilde{h}_{x}(f) = -i\sin\iota k_{\text{R}}^{-1/3}\tilde{g}_{\text{R}}^{(2)}(f)-\frac{4}{3}i\mathcal{G}k_{\text{R}}^{-1/6}\tilde{g}_{\text{R}}^{(1)}(f) \\
	&\tilde{h}_{y}(f) = -\frac{\sin{2\iota}}{2}k_{\text{R}}^{-1/3}\tilde{g}_{\text{R}}^{(2)}(f)-\frac{4}{3}\mathcal{G}\cos\iota k_{\text{R}}^{-1/6}\tilde{g}_{\text{R}}^{(1)}(f)
\end{split}
\end{equation}
where
\begin{equation}
	\tilde{g}_{\text{R}}^{(2)}(f) = k_{\text{R}}^{-5/12}i\sqrt{\frac{5\pi}{84}}\frac{\mathcal{M}^{2}}{D}(\pi\mathcal{M}f)^{-7/6}e^{-i\Psi_{\text{R}}^{(2)}(f)}
\end{equation}
and
\begin{equation}
	\tilde{g}_{\text{R}}^{(1)}(f) = k_{\text{R}}^{-5/12}i\sqrt{\frac{5\pi}{336}}\eta^{1/5}\frac{\mathcal{M}^{2}}{D}(\pi\mathcal{M}f)^{-3/2}e^{-i\Psi_{\text{R}}^{(1)}(f)}\,.
\end{equation}
The equations are presented here for giving readers the intuition of the injections that we choose, and we refer readers to the paper \cite{Chatziioannou_2012} for the definitions of the parameters. One could see that the dipole contribution $\tilde{g}_{R}^{(1)}(f)$ only enters the non-tensorial polarization modes. The strength of the dipole radiation mainly depends on the parameter $\mathcal{G}$ which is the difference in the self-gravitational binding energy per unit mass (or the difference in sensitivity \cite{1977ApJ...212L..91W,1989ApJ...346..366W}) of the binary components. We notice that in the paper \cite{Chatziioannou_2012} the definitions of $s_{1}$ and $s_{2}$ are not consistent between the expressions $\mathcal{G} = s_{1}/m_{1} - s_{2}/m_{2}$ and $k_{R} = 1-4s_{1}s_{2}/3$. The former is the self-gravitational binding energy \cite{1977ApJ...214..826W} but the latter is the self-gravitational binding energy per unit mass \cite{1977ApJ...212L..91W}. Here we adopt the latter definition, and therefore we have $\mathcal{G}=s_{1}-s_{2}$.

\begin{figure}
	\includegraphics[width=\linewidth]{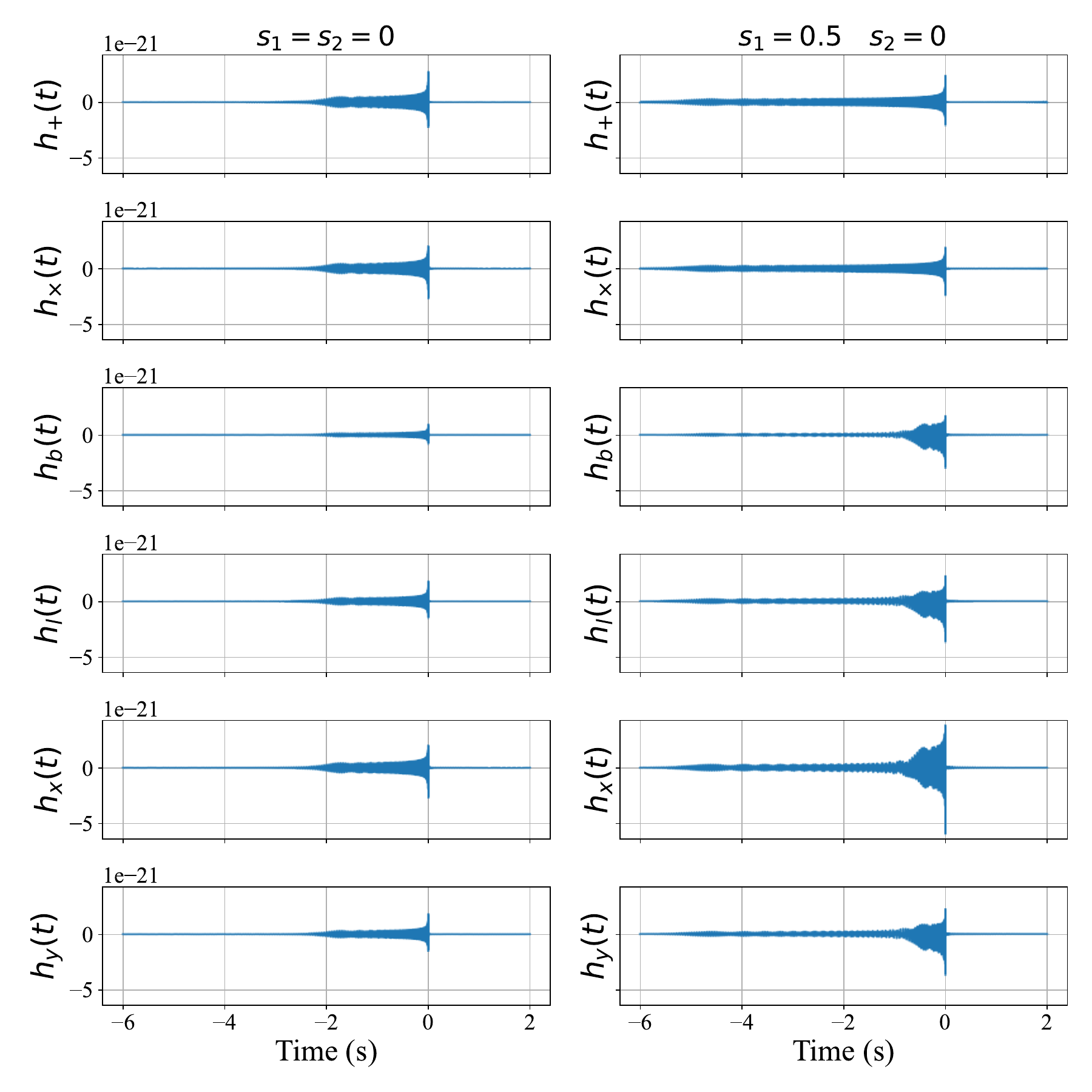}
	\caption{The plots show the polarization modes of the Rosen waveform. The component masses are $m_{1}=m_{2}=20M_{\odot}$. The coalescence phase is $0$. The inclination angle is $\pi/4$. The left panel shows the waveforms with sensitivities $s_{1} = s_{2} = 0$. The right panel shows the waveforms with sensitivities $s_{1} = 0.5$ and $s_{2} = 0$}
	\label{fig:result_rosen_waveform}
\end{figure}

\begin{figure*}
	\includegraphics[width=\linewidth]{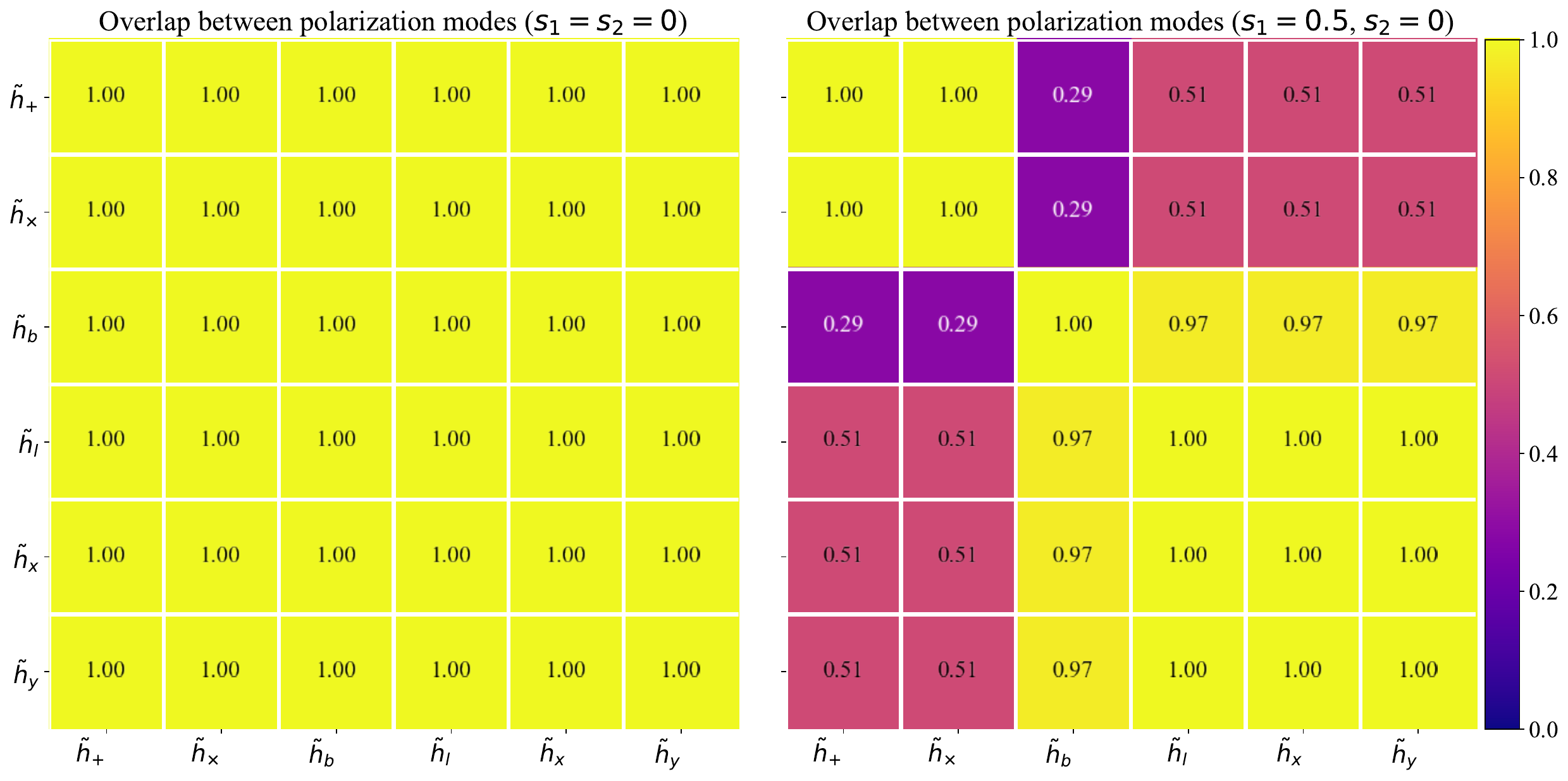}
	\caption{The left panel shows the overlap between the polarization modes when sensitivities are $s_{1} = s_{2} = 0$. Dipole radiation is not excited, and the polarization modes are linearly dependent with each other. The right panel shows the overlap between the polarization modes when the sensitivities are $s_{1} = 0.5$ and $s_{2} = 0$. Dipole radiation is excited, and therefore the polarization modes cannot be described with a single basis mode.}
	\label{fig:result_overlap}	
\end{figure*}

Waveforms are generated with a sampling rate $f_{s}=2048\text{ Hz}$ and a frequency lower cut $f_{\text{low}}=20\text{ Hz}$. The component masses are set to $m_{1}=m_{2}=20M_{\odot}$. The inclination angle is set to $\pi/4$. The coalescence phase and polarization angle are set to $0$. The geocentric GPS time is set to $1282107824$. The right ascension and declination of the source location are set to $\alpha=2.72$ and $\delta=-0.36$ respectively. We experiment on two different cases when the two compact objects have the same sensitivities i.e.\ $s_{1}=s_{2}$ and when they have very different sensitivities i.e.\ $s_{1}\neq s_{2}$. In the former case, the dipole radiation is not excited, and the polarization modes are well described with a single basis mode. In the latter case, the dipole radiation is excited, and the tensorial modes and non-tensorial modes cannot be well described with a single basis mode. We could compute the overlap $O(\tilde{h}_{a},\tilde{h}_{b})$ between two polarization modes $\tilde{h}_{a}(f)$ and $\tilde{h}_{b}(f)$ defined by
\begin{equation}
	O(\tilde{h}_{a},\tilde{h}_{b}) = \left|\frac{\int_{0}^{\infty}\tilde{h}_{a}^{*}\tilde{h}_{b}(f)df}{\sqrt{\int_{0}^{\infty}\left|\tilde{h}_{a}(f)\right|^{2}df\int_{0}^{\infty}\left|\tilde{h}_{b}(f)\right|^{2}df}}\right|
\end{equation}
to quantify how well the two polarization modes can be described with a single basis mode. The left panels of Fig.~\ref{fig:result_rosen_waveform} and Fig.~\ref{fig:result_overlap} show the time domain polarization modes with $s_{1}=s_{2}=0$ and the overlap between the polarization modes respectively. The right panel of Fig.~\ref{fig:result_rosen_waveform} and Fig.~\ref{fig:result_overlap} show the time domain polarization modes with $s_{1}=0.5$ and $s_{2}=0$ and the overlap between the polarization modes respectively. The polarization modes are projected onto the HLV network and injected into simulated Gaussian noise with the advanced LIGO and advanced Virgo design sensitivities i.e.\ \texttt{aLIGODesignSensitivityP1200087} and \texttt{AdVDesignSensitivityP1200087} \cite{Abbott_2020} respectively. Similar to the study with the ad hoc injections in Sec.~\ref{sec:ad_hoc_injections}, the injected signals are scaled to 20 different network SNRs equally spaced between 10 and 100. The power of each polarization mode relative to the $+$ polarization defined by
\begin{equation}
	E^{\text{rel}}_{m} = \frac{\int_{0}^{\infty}\left|\tilde{h}_{m}(f)\right|^{2}df}{\int_{0}^{\infty}\left|\tilde{h}_{+}(f)\right|^{2}df}
\end{equation}
where $\tilde{h}_{m}(f)$ is the polarization mode $m\in\{+,\times,b,l,x,y\}$ in the frequency domain is summarized in Table~\ref{tab:rosen_relative_energy}.
\begin{table}
	\begin{tabular}{|| c | c | c ||}
		\hline
		& $s_{1}=s_{2}=0$ & $s_{1}=0.5$, $s_{2}=0$ \\
		\hline
		Polarization mode & Relative power & Relative power \\ [0.5ex] 
		\hline\hline
		$\tilde{h}_{+}(f)$ & $1.00$ & $1.00$ \\
		\hline
		$\tilde{h}_{\times}(f)$ & $0.89$ & $0.89$ \\
		\hline
		$\tilde{h}_{b}(f)$ & $0.11$ & $1.35$ \\
		\hline
		$\tilde{h}_{l}(f)$ & $0.44$ & $1.68$ \\
		\hline
		$\tilde{h}_{x}(f)$ & $0.89$ & $3.36$ \\
		\hline
		$\tilde{h}_{y}(f)$ & $0.44$ & $1.68$ \\
		\hline
	\end{tabular}
	\caption{The table summarizes the relative power of the polarization modes with respect to the $+$ polarization mode of the Rosen waveform injections.}
	\label{tab:rosen_relative_energy}
\end{table}

\subsubsection{Effect of the orthogonal polarization component}
\label{sec:result_rosen_discussion}

\begin{figure}
	\includegraphics[width=\linewidth]{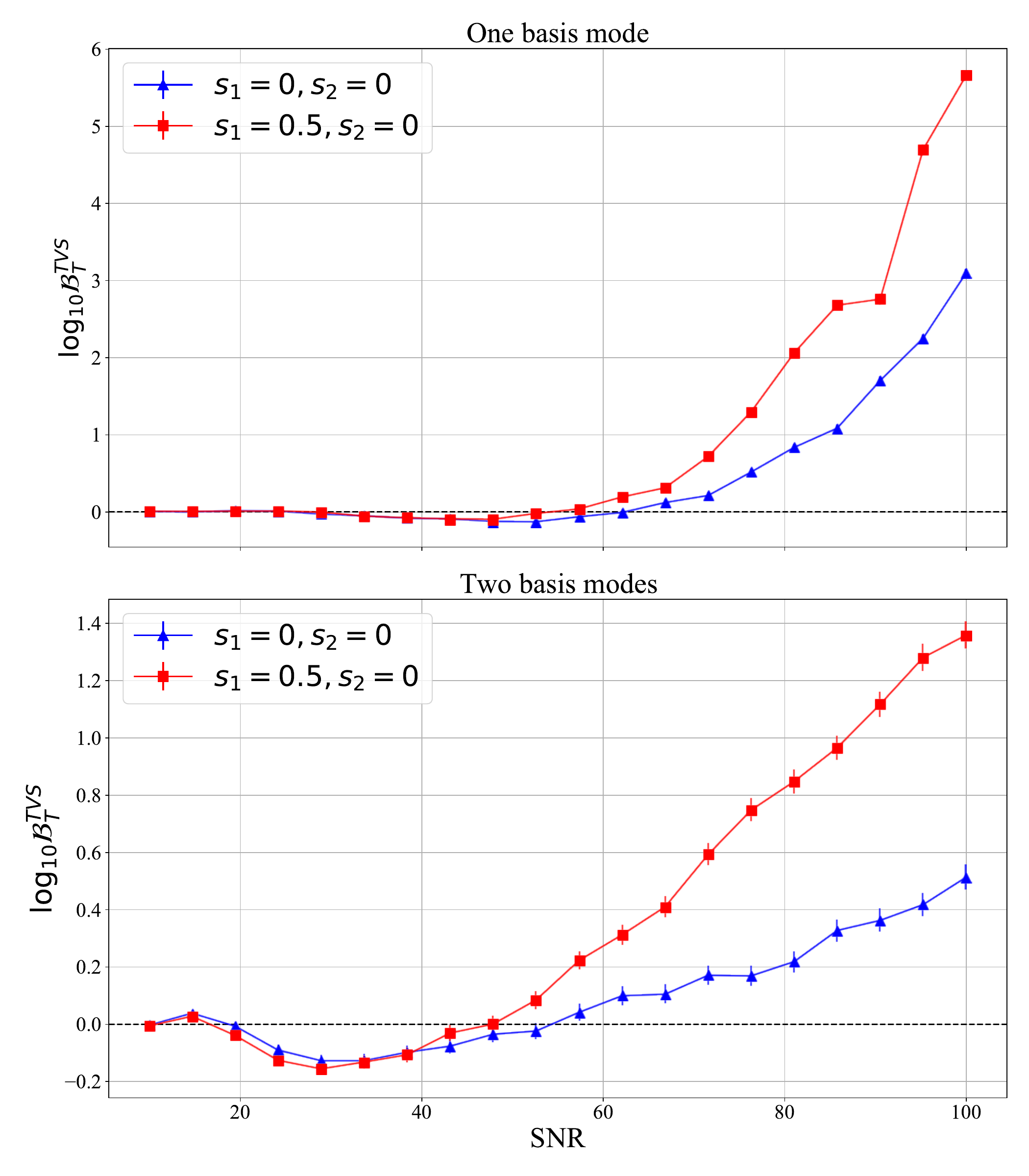}
	\caption{The plots show the variation of $\log_{10}\mathcal{B}_{T}^{TVS}$ with varying SNR. The upper panel shows the results with one basis mode. $+$ mode is chosen to be the basis mode for both $\mathcal{H}_{T}$ and $\mathcal{H}_{TVS}$. The beam pattern matrix of the one-basis-mode analysis is defined in Eq.~\eqref{eq:L1_pol}. The lower panel shows the results with two basis modes. $+$ mode and $\times$ mode are the basis modes for $\mathcal{H}_{T}$ which is also the only choice. $+$ mode and vector $x$ mode are chosen to be the basis modes for $\mathcal{H}_{TVS}$. The beam pattern matrix of the two-basis-mode analysis is defined in Eq.~\eqref{eq:L2_pol}. The triangles denote the results with the injections of equal sensitivities ($s_{1}=s_{2}=0$), and the squares denote the results with the injections of unequal sensitivities ($s_{1}=0.5$ and $s_{2}=0$). The error bars in the upper panel are too small to be seen.}
	\label{fig:result_rosen_logB}
\end{figure}

To begin with, we first discuss how would the orthogonal polarization component affect the detection. To quantify the linear dependency between the polarization modes, we may compute the effective rank \cite{effective_rank} of the matrix of polarization modes
\begin{equation}
	\tilde{\boldsymbol{h}}=
	\begin{bmatrix}
		\tilde{h}_{+}[1] & \tilde{h}_{\times}[1] & \tilde{h}_{b}[1] & \tilde{h}_{l}[1] & \tilde{h}_{x}[1] & \tilde{h}_{y}[1] \\
		\tilde{h}_{+}[2] & \tilde{h}_{\times}[2] & \tilde{h}_{b}[2] & \tilde{h}_{l}[2] & \tilde{h}_{x}[2] & \tilde{h}_{y}[2] \\
		\vdots & \vdots & \vdots & \vdots & \vdots & \vdots \\
		\tilde{h}_{+}[K] & \tilde{h}_{\times}[K] & \tilde{h}_{b}[K] & \tilde{h}_{l}[K] & \tilde{h}_{x}[K] & \tilde{h}_{y}[K] \\
	\end{bmatrix}
\end{equation}
where $K$ is the number of frequency bins. The effective rank for the equal sensitivity injection shown in the left panel of Fig.~\ref{fig:result_rosen_waveform} is $1$ which is obvious which implies the polarization modes can be well described with a single basis mode. The effective rank for the unequal sensitivity injection shown in the right panel of Fig.~\ref{fig:result_rosen_waveform} is $1.83$ which implies the polarization modes cannot be well described with a single basis mode. The results of the analysis are shown in Fig.~\ref{fig:result_rosen_logB}. The upper panel shows the results using one basis mode. $+$ mode is chosen to be the basis mode for both $\mathcal{H}_{T}$ and $\mathcal{H}_{TVS}$. The lower panel shows the results using two basis modes. $+$ mode and $\times$ mode are chosen to be the basis modes for $\mathcal{H}_{T}$ which is also the only choice. $+$ mode and vector $x$ mode are chosen to be the basis modes for $\mathcal{H}_{TVS}$. The blue dots show the results with the injections of equal sensitivities. In this case the dipole radiation is not excited and there is no orthogonal polarizations as shown in Eq.~\eqref{eq:rosen_waveform} and the left panel of Fig.~\ref{fig:result_rosen_waveform}. The red dots show the results with the injections of unequal sensitivities. As shown in Eq.~\eqref{eq:rosen_waveform} and the right panel of Fig.~\ref{fig:result_rosen_waveform}, the polarization modes can not be well described with a single basis mode. Nevertheless, the red dots in the upper panel still show a general trend favoring $\mathcal{H}_{TVS}$ even when the orthogonal polarization component is significant. A much stronger model preference is observed when the dipole radiation presents in the signal in the $L=1$ analysis even in this case the orthogonal polarization component is strong.

\begin{figure}
	\includegraphics[width=\linewidth]{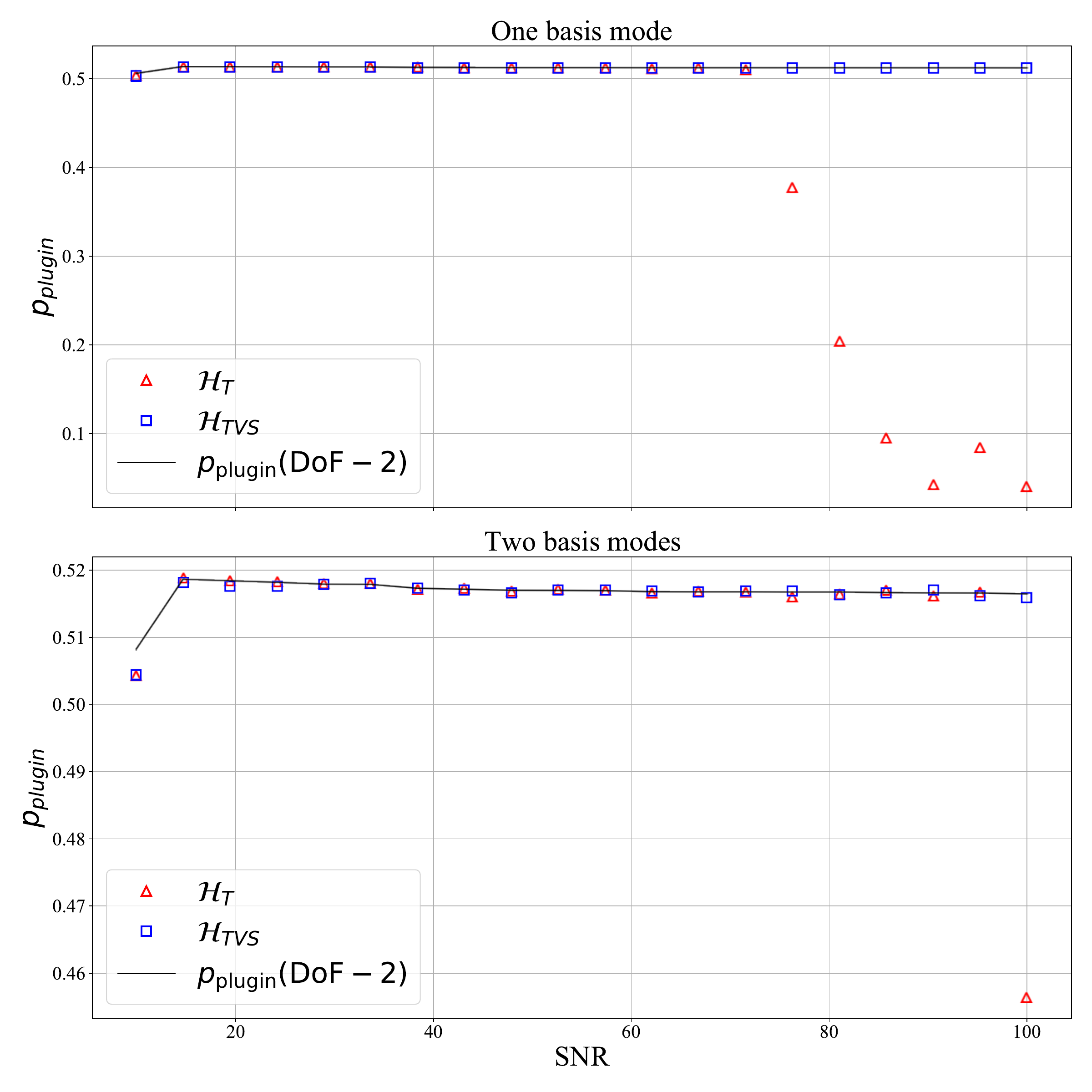}
	\caption{The plot shows the plug-in p-value $p_{\text{plugin}}$ with the injections of equal sensitivities with different SNRs. The upper panel shows the variation of $p_{\text{plugin}}$ of the analyses with one basis mode. The bases of $\mathcal{H}_{T}$ and $\mathcal{H}_{TVS}$ are both chosen to be the $+$ mode. The beam pattern matrix of the one-basis-mode analysis is defined in Eq.~\eqref{eq:L1_pol}. The lower panel shows the variation of $p_{\text{plugin}}$ of the analyses with two basis modes. The basis of $\mathcal{H}_{T}$ is $+$ mode and $\times$ mode. The basis of $\mathcal{H}_{TVS}$ is chosen to be $+$ mode and vector $x$ mode. The beam pattern matrix of the two-basis-mode analysis is defined in Eq.~\eqref{eq:L2_pol}. The triangles denote the plug-in p-value of the analysis with $\mathcal{H}_{T}$. The squares denote the plug-in p-value of the analysis with $\mathcal{H}_{TVS}$. The solid line indicates the p-value of the residual power when it is exactly equal to the mode of the $\chi^{2}$ distribution.}
	\label{fig:result_p_val_equal_s}
\end{figure}

\begin{figure}
	\includegraphics[width=\linewidth]{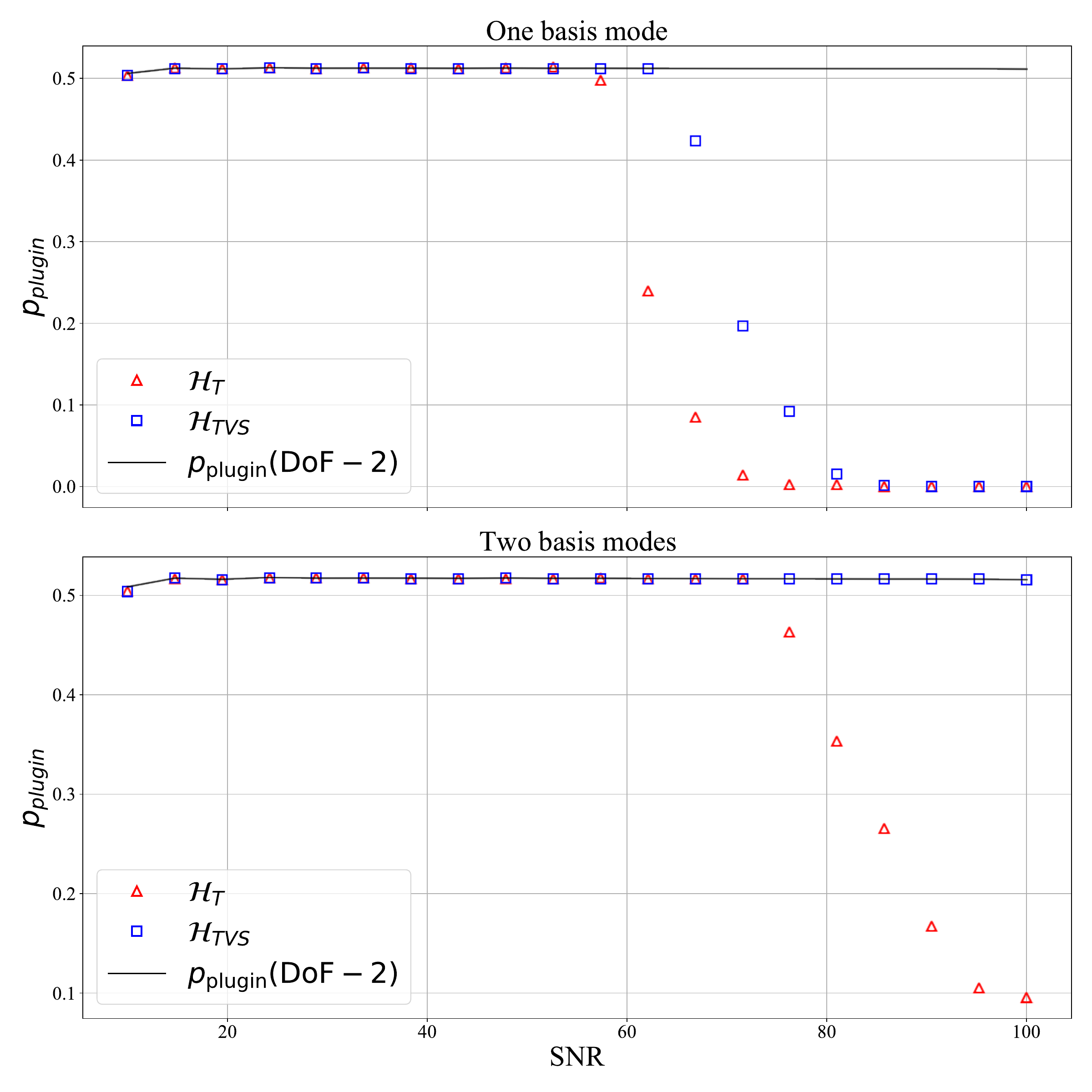}
	\caption{The plot shows the plug-in p-value $p_{\text{plugin}}$ with the injections of unequal sensitivities with different SNRs. The upper panel shows the variation of $p_{\text{plugin}}$ of the analyses with one basis mode. The bases of $\mathcal{H}_{T}$ and $\mathcal{H}_{TVS}$ are both chosen to be the $+$ mode. The beam pattern matrix of the one-basis-mode analysis is defined in Eq.~\eqref{eq:L1_pol}. The lower panel shows the variation of $p_{\text{plugin}}$ of the analyses with two basis modes. The basis of $\mathcal{H}_{T}$ is $+$ mode and $\times$ mode. The basis of $\mathcal{H}_{TVS}$ is chosen to be $+$ mode and vector $x$ mode. The beam pattern matrix of the two-basis-mode analysis is defined in Eq.~\eqref{eq:L2_pol}. The triangles denote the plug-in p-value of the analysis with $\mathcal{H}_{T}$. The squares denote the plug-in p-value of the analysis with $\mathcal{H}_{TVS}$. The solid line indicates the p-value of the residual power when it is exactly equal to the mode of the $\chi^{2}$ distribution.}
	\label{fig:result_p_val_unequal_s}
\end{figure}

Fig.~\ref{fig:result_p_val_equal_s} and Fig.~\ref{fig:result_p_val_unequal_s} show the plug-in p-value of the injections as discussed in Sec.~\ref{sec:orthogonal} to test the presence of uncaptured orthogonal polarizations. The black line labeled as $p_{\text{plugin}}(\text{DoF}-2)$ in the figures is reference p-value when the null energy attains the highest possible likelihood.

Fig.~\ref{fig:result_p_val_equal_s} shows the plug-in p-value of the injections with equal sensitivities. In the upper panel, we could see that the plug-in p-values of $\mathcal{H}_{TVS}$ are consistent with the black line for all SNRs. It implies that there always exists a null operator constructed from a linear combination of $\boldsymbol{f}_{\{+,\times,b,l,x,y\}}$ with one column to produce a residual that is consistent with noise, and no orthogonal polarization component is observed. And indeed the injected polarization modes can be well described with a single basis mode. But for the red dots that correspond to $\mathcal{H}_{T}$, the plug-in p-value drops suddenly when the SNR increases through $\sim75$. This is expected since the injected signals carry tensor, vector, and scalar components, and a null operator constructed from $\boldsymbol{f}_{\{+,x\}}$ could not completely cancel the signal, but such insufficiency only shows up with a high enough SNR.

Fig.~\ref{fig:result_p_val_unequal_s} shows the plug-in p-value of the injections with unequal sensitivities. The upper panel shows the results with one basis mode. The plug-in p-value decreases with a higher SNR for both $\mathcal{H}_{TVS}$ and $\mathcal{H}_{T}$. This is expected since the injected signals carry a strong dipole component, and we could not explain the data well with a single basis mode. This suggests the feasibility to use the plug-in p-value as a tool to diagnose the validity of the assumed number of basis modes. A low plug-in p-value also gives us an alarming message if the $\tilde{h}_{+}(f)$ and $\tilde{h}_{\times}(f)$ in the GR waveform model can be well described with a single basis mode, and this would indicate we have observed something that cannot be explained by the GR waveform model. The lower panel shows the results using two basis modes. For the $\mathcal{H}_{TVS}$, we choose the $+$ mode and the vector $x$ mode as the basis. The plug-in p-values are consistent with $p_{\text{plugin}}(\text{DoF}-2)$ for all SNRs. As shown in the right panel of Fig.~\ref{fig:result_overlap}, we could choose any one of the tensorial modes and any one of the non-tensorial modes to construct the basis, and the null operator constructed from the basis would be sufficient to cancel the signal. On the other hand, the plug-in p-value for $\mathcal{H}_{T}$ drops when the SNR is sufficiently high. Even when we allow $\tilde{h}_{+}$ and $\tilde{h}_{\times}$ to take arbitrary forms independently, the null operator could not cancel the signal and it results in an excess amount of residual power.

The example shows that the method is capable of capturing the parallel non-tensorial polarization component, and the presence of a significant orthogonal polarization component would not deteriorate the method. We have also demonstrated using the plug-in p-value to test the existence of the orthogonal polarization component.

\subsubsection{Ranking among polarization hypotheses}
\label{sec:pol_rank}
We also perform the analysis with different polarization hypotheses to investigate how would the pipeline rank different hypotheses. The choices of basis and the dimensionality of the parameter space are summarized in Table~\ref{tab:rosen_param_sum}.
\begin{table*}
	\begin{tabular}{|| c | c | c | c | c ||}
		\hline
		Hypothesis & Description & Mode(s) & Basis mode(s) & Number of parameters \\ [0.5ex] 
		\hline\hline
		$\mathcal{H}_{T,1}$ & Pure tensorial & $+$, $\times$ & $+$ & 5  \\
		\hline
		$\mathcal{H}_{V,1}$ & Pure vectorial & $x$, $y$ & $x$ & 5  \\
		\hline
		$\mathcal{H}_{S,1}$ & Pure scalar & $b$ & $b$ & 2  \\
		\hline
		$\mathcal{H}_{TS,1}$ & Tensor-scalar & $+$, $\times$, $b$, $l$ & $+$ & 9 \\
		\hline
		$\mathcal{H}_{TV,1}$ & Tensor-vector & $+$, $\times$, $x$, $y$ & $+$ & 9 \\
		\hline
		$\mathcal{H}_{VS,1}$ & Vector-scalar & $x$, $y$, $b$, $l$ & $x$ & 9 \\
		\hline
		$\mathcal{H}_{TVS,1}$ & Tensor-vector-scalar & $+$, $\times$, $x$, $y$, $b$, $l$ & $+$ & 13  \\
		\hline
		$\mathcal{H}_{T,2}$ & Pure tensorial & $+$, $\times$ & $+$, $\times$ & 2 \\
		\hline
		$\mathcal{H}_{V,2}$ & Pure vectorial & $x$, $y$ & $x$, $y$ & 2 \\
		\hline
		$\mathcal{H}_{TS,2}$ & Tensor-scalar & $+$, $\times$, $b$, $l$ & $+$, $b$ & 11 \\
		\hline
		$\mathcal{H}_{TV,2}$ & Tensor-vector & $+$, $\times$, $x$, $y$ & $+$, $x$ & 11 \\
		\hline
		$\mathcal{H}_{VS,2}$ & Vector-scalar & $x$, $y$, $b$, $l$ & $x$, $b$ & 11 \\
		\hline
		$\mathcal{H}_{TVS,2}$ & Tensor-vector-scalar & $+$, $\times$, $x$, $y$, $b$, $l$ & $+$, $x$ & 19 \\
		\hline
	\end{tabular}
	\caption{The table summarizes the choice of basis used in the analysis discussed in Sec.~\ref{sec:pol_rank} and the number of model parameters of each polarization hypothesis.}
	\label{tab:rosen_param_sum}
\end{table*}

\begin{figure*}
	\includegraphics[width=\linewidth]{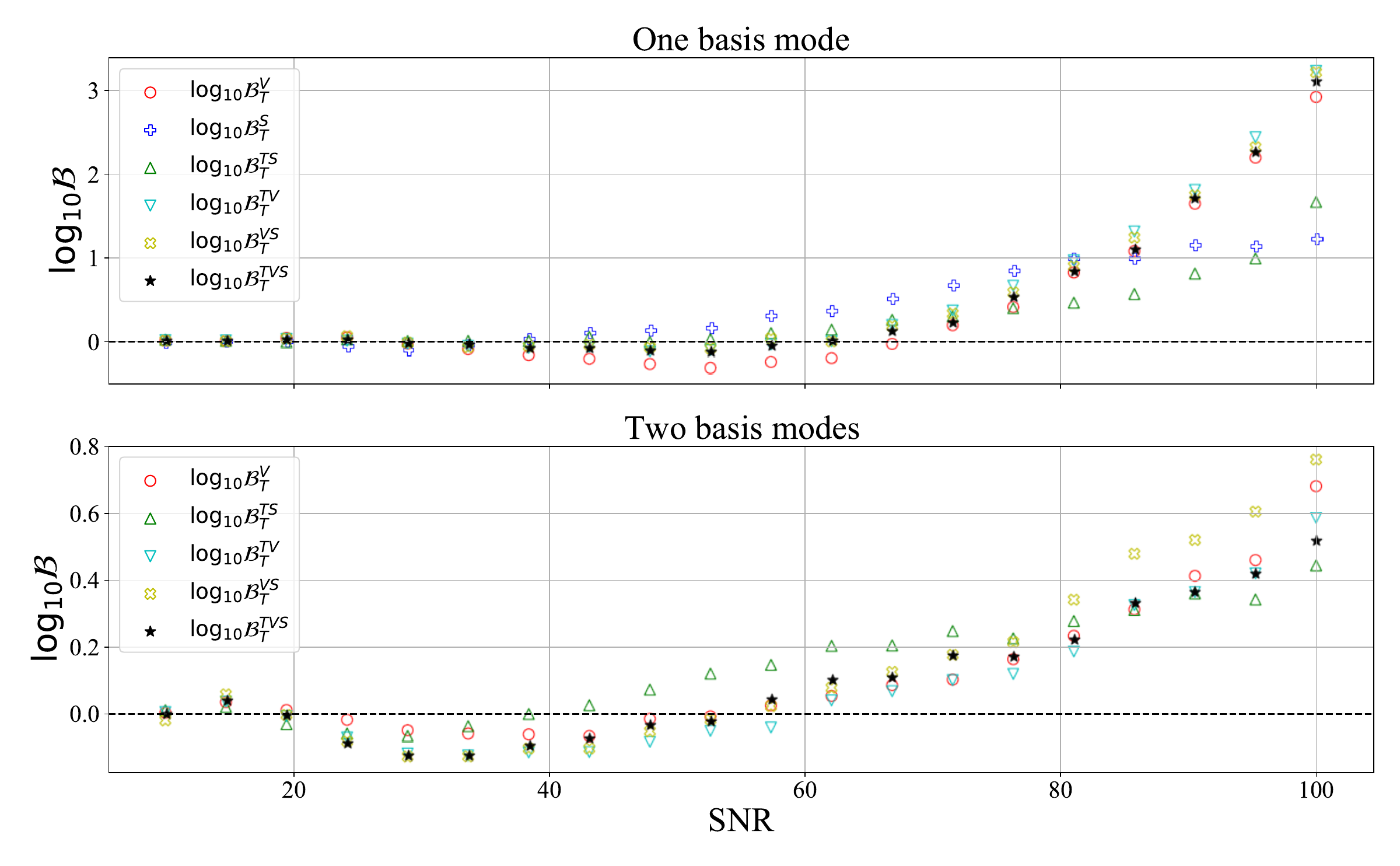}
	\caption{The plot shows the variation of $\log_{10}$ Bayes factor of different polarization hypotheses against the tensor hypothesis with the Rosen waveform injections with $s_{1}=s_{2}=0$ and different SNRs. The upper panel shows the results with one basis mode. The beam pattern matrix of the one-basis-mode analysis is defined in Eq.~\eqref{eq:L1_pol}. The lower panel shows the results with two basis modes. The beam pattern matrix of the two-basis-mode analysis is defined in Eq.~\eqref{eq:L2_pol}.}
	\label{fig:result_rosen_ranking_equal_s}
\end{figure*}

\begin{figure}
	\includegraphics[width=\linewidth]{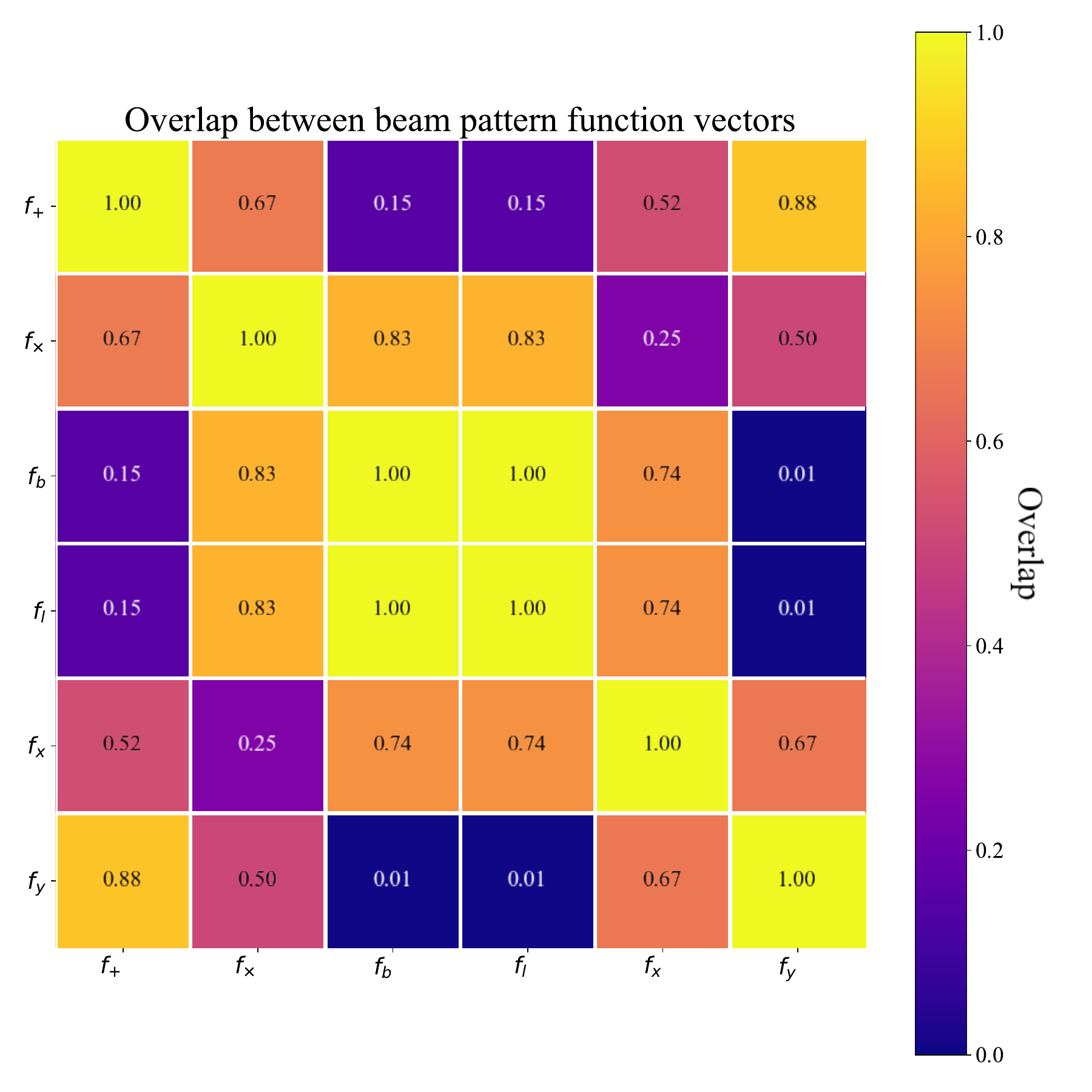}
	\caption{The figure shows the overlap between the beam pattern function vectors $\boldsymbol{f}_{m}$ with the injected source location i.e.\ $\alpha=2.72$ and $\delta=-0.36$, polarization angle $\psi=0$ and GPS time $1282107824$.}
	\label{fig:result_overlap_beam_pattern}
\end{figure}

Fig.~\ref{fig:result_rosen_ranking_equal_s} shows the result of the analyses of the injections with equal sensitivities i.e.\ $s_{1}=s_{2}=0$. The plot shows the variation of the $\log_{10}$ Bayes factor of different polarization hypotheses against the tensor hypothesis. As the SNR increases, the non-tensor hypotheses are more favored. The upper panel shows the result with one basis mode. Among the hypotheses, $\mathcal{H}_{V}$, $\mathcal{H}_{TV}$, $\mathcal{H}_{VS}$, and $\mathcal{H}_{TVS}$ are the most favored in the high SNR cases. When the SNR is between around $40$ and $80$, $\mathcal{H}_{S}$ is slightly more favored. This is partly because $\mathcal{H}_{S}$ involves the least number of parameters than other hypotheses as shown in Table~\ref{tab:rosen_param_sum} and therefore is the least penalized by Ockham's razor. The another more important reason is related to the overlap between the beam pattern function vectors defined by 
\begin{equation}
	\label{eq:beam_overlap}
	\mathcal{D}(\hat{\Omega};a,b) = \left|\frac{\boldsymbol{f}_{a}(\hat{\Omega})\cdot\boldsymbol{f}_{b}(\hat{\Omega})}{\sqrt{\boldsymbol{f}_{a}(\hat{\Omega})\cdot\boldsymbol{f}_{a}(\hat{\Omega})}\sqrt{\boldsymbol{f}_{b}(\hat{\Omega})\cdot\boldsymbol{f}_{b}(\hat{\Omega})}}\right|\in[0,1]\,.
\end{equation}
where $\hat{\Omega}$ is the sky location of the source, and $\boldsymbol{f}_{a}$ and $\boldsymbol{f}_{b}$ are the beam patten vectors of polarizations $a$ and $b$.
Fig.~\ref{fig:result_overlap_beam_pattern}. A higher overlap at the sky location $\hat{\Omega}$ implies it is more difficult to distinguish the polarizations $a$ and $b$ when the GW comes from that sky position. Fig.~\ref{fig:result_overlap_beam_pattern} shows the overlap between the beam pattern vectors at the injected source location. We could see that $\boldsymbol{f}_{b}$ has a very significant overlap with $\boldsymbol{f}_{\times}$ and $\boldsymbol{f}_{x}$. This implies that the signal power from $\tilde{h}_{\times}(f)$ and $\tilde{h}_{x}(f)$ can be significantly reduced with the null operator constructed from solely $\boldsymbol{f}_{b}$, and this causes some confusion with the other hypotheses. The lower panel shows the result using two basis modes. The $\mathcal{H}_{VS}$ is the most favored instead of $\mathcal{H}_{TVS}$ in the high SNR cases. This is again due to the overlap between the beam pattern vectors. One could see from Fig.~\ref{fig:result_overlap_beam_pattern} that $\boldsymbol{f}_{b}$ has an overlap $0.83$ with $\boldsymbol{f}_{\times}$ and $\boldsymbol{f}_{y}$ has an overlap $0.88$ with $\boldsymbol{f}_{+}$. One should be reminded that the polarization modes are linearly dependent when the sensitivities are equal, and therefore we could reduce a significant amount of power from the tensorial modes by the null operator constructed in $\mathcal{H}_{VS}$. Also, the much larger parameter space of $\mathcal{H}_{TVS}$ (19 parameters) compared with $\mathcal{H}_{VS}$ (11 parameters) adds a penalty to the evidence.

\begin{figure*}
	\includegraphics[width=\linewidth]{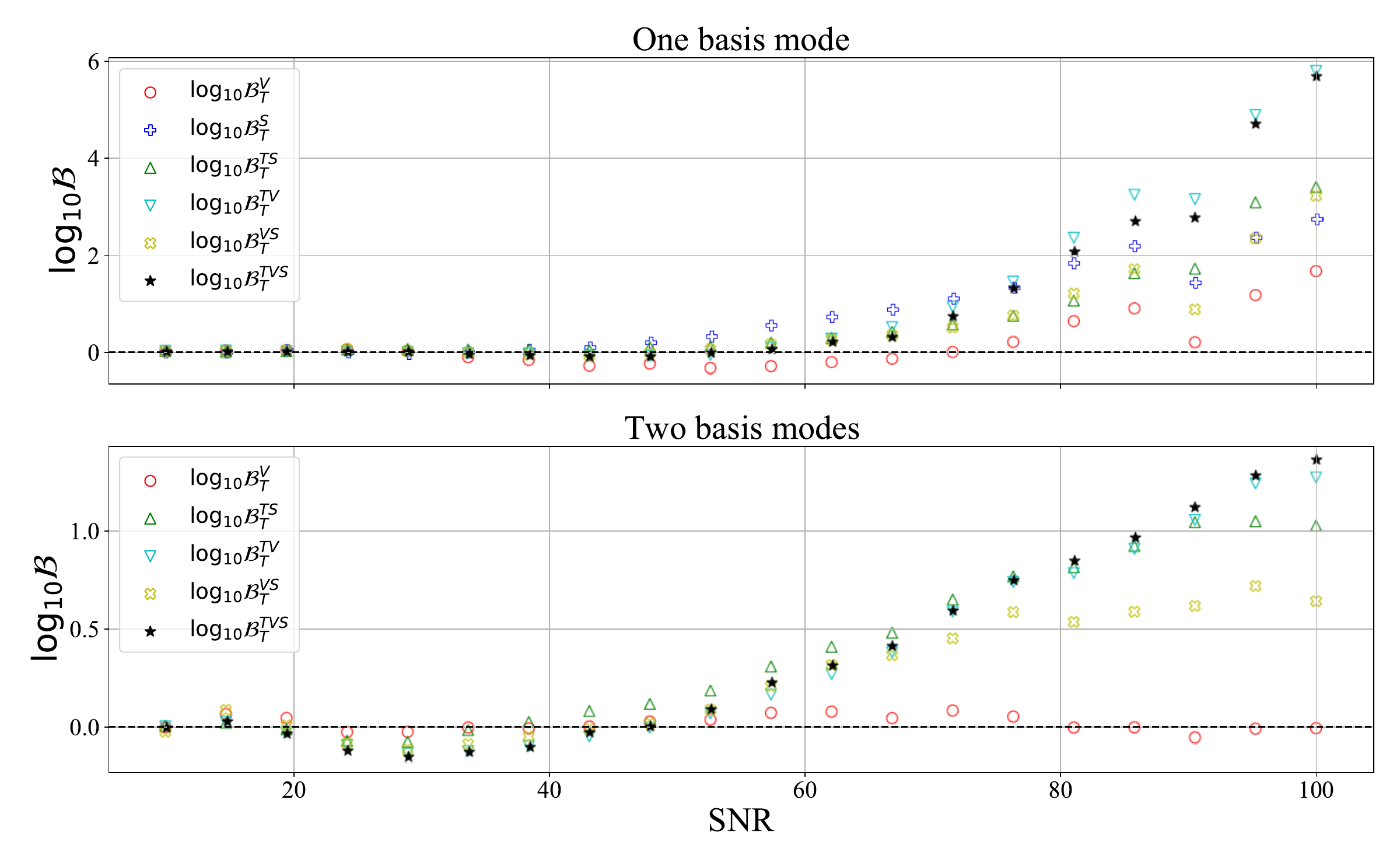}
	\caption{The plot shows the variation of $\log_{10}$ Bayes factor of different polarization hypotheses against the tensor hypothesis with the Rosen waveform injections with $s_{1}=0.5$, $s_{2}=0$ and different SNRs. The upper panel shows the results with one basis mode. The beam pattern matrix of the one-basis-mode analysis is defined in Eq.~\eqref{eq:L1_pol}. The lower panel shows the results with two basis modes. The beam pattern matrix of the two-basis-mode analysis is defined in Eq.~\eqref{eq:L2_pol}.}
	\label{fig:result_rosen_ranking_unequal_s}
\end{figure*}

Fig.~\ref{fig:result_rosen_ranking_unequal_s} shows the result of the analyses of the injections with different sensitivities i.e.\ $s_{1}=0.5$ and $s_{2}=0$. In this case, the dipole radiation is excited, and we have to use at least two basis modes to explain the data. The upper panel shows the result with one basis mode. Even when the orthogonal polarization component is significant, $\mathcal{H}_{TVS}$ still has a very high rank among the hypotheses. $\mathcal{H}_{TV}$ is slightly more favored due to the extra penalization on the larger parameter space of $\mathcal{H}_{TVS}$. The lower panel shows the result with two basis modes. In this case, there is no orthogonal polarization component remains with the best-fit parameters in $\mathcal{H}_{TVS}$ as shown in the lower panel of Fig.~\ref{fig:result_p_val_unequal_s}, and indeed $\mathcal{H}_{TVS}$ is more favored over $\mathcal{H}_{TV}$ in the high SNR cases even though it has a much larger parameter space.

\subsubsection{Discussion}

The results in Fig.~\ref{fig:result_rosen_ranking_unequal_s} are in agreement with our discussion in Sec.~\ref{sec:orthogonal}. The existence of a significant orthogonal polarization component does not deteriorate the capability of the method to detect a GR violation as shown in the upper panel of the figure. Similar to the results in Sec.~\ref{sec:ad_hoc_injections}, one should not overinterpret the figures to be stating the expected deviations that we would observe since the source location of all injections is arbitrarily set to $\alpha=2.72$ and $\delta=-0.36$, but the distinguishability between different polarization hypotheses significantly depends on the source location. We have shown that confusion between polarization hypotheses can arise when there is a significant overlap between the beam pattern vectors. In terms of identifying the true polarization hypothesis, the most important factor is the overlap between the beam pattern vectors that determines how well we can distinguish between different polarizations. The overlap can be significantly reduced by having a larger detector network in the future.

\section{Summary and conclusions}
\label{sec:conclusion}
We have presented a null-stream-based generic Bayesian unmodeled framework to probe GW polarizations. We proposed the basis formulation to reformulate the generic null projection along the beam pattern vectors to the null projection along the polarization basis modes. The advantage of this formulation is to guarantee the equal dimensionality of the residuals after performing the null projection for fair model comparison, and we do not need to explicitly assume the waveform of the basis modes. This gives the generality of the framework to probe GW polarizations without requiring modeling non-GR waveform explicitly. This method would be useful when the non-GR waveforms are not well developed. 

We first conducted a mock data study with the ad hoc injections and perform the one-basis-mode ($L=1$) analysis and the two-basis-mode ($L=2$) analysis. The ad hoc injections are generated by projecting the plus polarization and the cross polarization to the non-tensorial beam pattern functions. In this case, the polarization modes can be well described with a single basis mode. This serves as a sanity check to investigate the capability of the $L=1$ analysis and the $L=2$ analysis to detect GW polarizations when there is no uncaptured orthogonal polarization component. We performed the mock data study with tensor, vector, scalar, tensor-scalar, tensor-vector, vector-scalar, and tensor-vector-scalar injections in the HLV 3-detector network at the design sensitivity. The source location is not known and we marginalize over the whole sky sphere in the analysis. We varied the signal-to-noise ratio (SNR) of the injections and we showed that the non-tensor hypotheses are more favored with an increasing SNR and increasing strength of non-tensorial components of the injections. We then conducted a mock data study with the more realistic non-GR waveform i.e. the inspiral waveforms of Rosen's theory that predicts the existence of all six polarization states. We considered two different cases in which the dipole radiation is excited and not excited respectively. The presence of strong dipole radiation would give rise to a significant orthogonal polarization component in the $L=1$ analysis. Nevertheless, the $L=1$ analysis significantly favors $\mathcal{H}_{TVS}$ over $\mathcal{H}_{T}$. We also demonstrated the feasibility of using the plug-in p-value to test the presence of an orthogonal polarization component.

Lastly, we investigated how the pipeline ranks different polarization hypotheses with the Rosen waveform injections with equal and unequal sensitivities respectively. We showed that the presence of the orthogonal polarization component contributed by the dipole radiation does not deteriorate the $L=1$ analysis, and the true polarization hypothesis is one of the top-ranked hypotheses. We showed that there exists some confusion between different polarization hypotheses due to the penalty on the more complicated polarization model and the overlap between the beam pattern vectors. The polarization subspaces spanned by different beam pattern vectors are in general not orthogonal to each other, and the overlap is arguably the most important factor to distinguish between different polarization hypotheses. The overlap depends significantly on the sky location of the source, and in general, it can be reduced with a larger detector network. We should expect to constrain the GW polarization content a lot better in the future by including KAGRA \cite{akutsu2020overview,PhysRevD.88.043007}, LIGO India \cite{ligo_india}, Einstein telescope \cite{Punturo_2010,Hild_2011}, and Cosmic Explorer \cite{Abbott_2017_cosmic,reitze2019cosmic} in the joint observing runs.

Our work also demonstrated the feasibility to search for mixed polarizations with a limited number of detectors. Although we need at least $M+1$ non-coaligned detectors to reconstruct $M$ independent polarization modes, it is possible to detect the presence of extra polarization modes with a fewer number of detectors. We emphasize that this conclusion does not only apply to null-stream-based methods, and it should also motivate other approaches to search for mixed polarizations in the current LIGO-Virgo 3-detector network.
\section{Acknowledgement}

The authors would like to thank Giancarlo Cella for the comments in preparing the paper. ICFW and TGFL are partially supported by grants from the Research Grants Council of the Hong Kong (Project No. 24304317 and 14306419) and Research Committee of the Chinese University of Hong Kong. PTHP and CVDB are supported by the research program of the Netherlands Organization for Scientific Research (NWO). RKLL and TGFL would also like to gratefully acknowledge the support from the Croucher Foundation in Hong Kong. We are grateful for computational resources provided by Cardiff University, and funded by an STFC grant supporting UK Involvement in the Operation of Advanced LIGO. We are grateful for computational resources provided by the Leonard E Parker Center for Gravitation, Cosmology and Astrophysics at the University of Wisconsin-Milwaukee. We acknowledge the use of IUCAA LDG cluster Sarathi for the computational/numerical work.

\textit{Softwares}: \texttt{Eigen} \cite{eigenweb}, \texttt{FFTW3} \cite{1386650}, \texttt{GSL} \cite{10.5555/1538674}, \texttt{HDF5} \cite{hdf5}, \texttt{HEALPix} \cite{Gorski_2005}, \texttt{LALSuite} \cite{lalsuite} and \texttt{MultiNest} \cite{Feroz_2008,Feroz_2009,Feroz_2019} are used to perform analyses. Plots are generated using \texttt{Matplotlib} \cite{4160265} and \texttt{NumPy} \cite{harris2020array}.
\appendix

\section{Discrete Fourier transform}
\label{app:DFT}
In this section, we decribe the convention of discrete Fourier transform that we use. Given a discrete-time time series $x[n]$ of length $N$, the discrete Fourier transform $\tilde{x}[k]$ is defined as
\begin{equation}
	\tilde{x}[k]=\sum_{n=1}^{N}x[n]e^{-i2\pi nk/N}\Delta t
\end{equation}
where $\Delta t$ is the sampling interval, and $i^{2}=-1$. The inverse transform is defined as
\begin{equation}
	{x}[n]=\sum_{k=1}^{N}\tilde{x}[k]e^{i2\pi nk/N}\Delta f
\end{equation}
where $\Delta f=1/(N\Delta t)$ is the frequency resolution.

\section{Null stream}
\label{app:null_stream}

In this section, we discuss several properties of null stream including the reduced dimensionality of residuals and the role of polarization angle in the generic null stream construction.

\subsection{Reduced dimensionality}
\label{app:reduced_dim}

\begin{figure*}
	\includegraphics[width=0.8\linewidth]{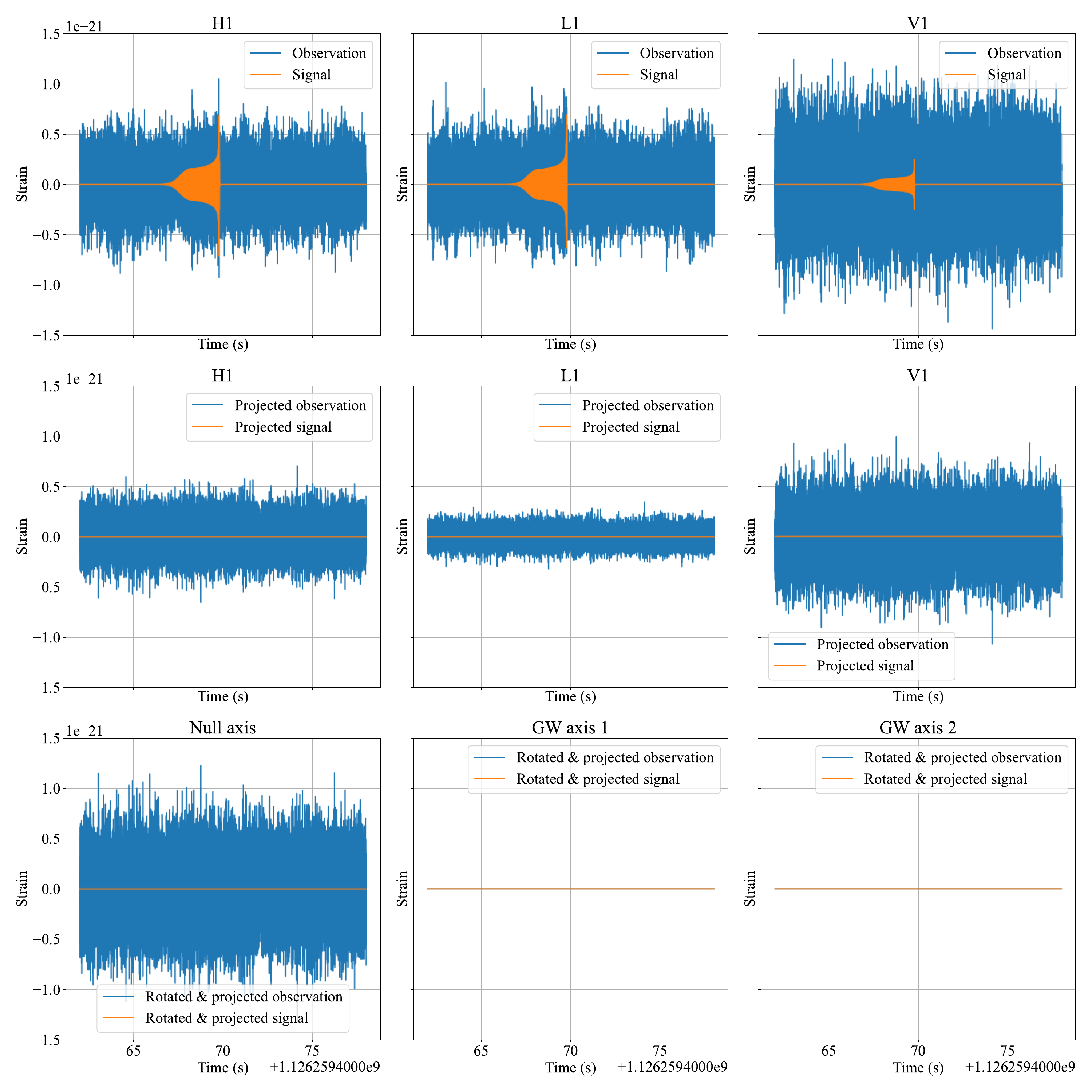}
	\caption{The figure shows an example of null stream construction. The upper panel shows the observed strain outputs and the pure tensorial signals in Hanford (H1), Livingston (L1), and Virgo (V1) respectively. The middle panel shows the residual in each detector after applying the correct null operator that is the null stream. The lower panel shows the null stream in the principal coordinate system. The principal axes are the two so-called GW axes on the hyperplane spanned by $\boldsymbol{f}_{+}$ and $\boldsymbol{f}_{\times}$. The null axis is orthogonal to the GW axes and no tensorial GW signal presents in this dimension.}
	\label{fig:null_stream_demo}
\end{figure*}

Null projection removes all data on the hyperplane spanned by the constituent beam pattern function vectors $\boldsymbol{f}_{m}$ used in the construction of the null operator. The resultant null stream, therefore, has a lower dimensionality than the original data. Fig.~\ref{fig:null_stream_demo} demonstrates an example of null projection. For the sake of a more intuitive demonstration, the data are displayed in the time domain. One should notice that in the middle panel the signal content is removed and the amplitude of the residual is reduced compared to the strain outputs. This is due to the reduced dimensionality after applying the null operator. The reduced dimensionality could be observed more readily by orienting the residuals along the principal axes. Recall Eq.~\eqref{eq:residual_start}, the null projection is
\begin{equation}
	\tilde{\boldsymbol{z}} := (\boldsymbol{I}-\boldsymbol{F}_{w}(\boldsymbol{F}_{w}^{\dagger}\boldsymbol{F}_{w})^{-1}\boldsymbol{F}_{w}^{\dagger})\boldsymbol{\mathcal{T}}\left(\tilde{\boldsymbol{d}}_{w};\Delta\boldsymbol{t}\right)\,.
\end{equation}
We perform the singular value decomposition of each frequency component of $\boldsymbol{I}-\boldsymbol{F}_{w}(\boldsymbol{F}_{w}^{\dagger}\boldsymbol{F}_{w})^{-1}\boldsymbol{F}_{w}^{\dagger}$.
\begin{equation}
	(\boldsymbol{I}-\boldsymbol{F}_{w}(\boldsymbol{F}_{w}^{\dagger}\boldsymbol{F}_{w})^{-1}\boldsymbol{F}_{w}^{\dagger})[k] = \boldsymbol{U}_{k}\boldsymbol{\Sigma}_{k}\boldsymbol{V}_{k}^{\dagger}
\end{equation}
where $\boldsymbol{U}\in\mathbb{C}^{D\times D}$ is a unitary matrix, $\boldsymbol{\Sigma}_{k}\in\mathbb{R}^{D\times D}$ is a diagonal matrix with the singular values as the diagonal entries, $\boldsymbol{V}_{k}\in\mathbb{C}^{D\times D}$ is a unitary matrix, and the subscript $k$ denotes the decomposition of the $k^{\text{th}}$ frequency component of the projector $\boldsymbol{I}-\boldsymbol{F}_{w}(\boldsymbol{F}_{w}^{\dagger}\boldsymbol{F}_{w})^{-1}\boldsymbol{F}_{w}^{\dagger}$. The null stream in the principal coordinate system is obtained by applying the rotation matrix $\boldsymbol{U}_{k}^{\dagger}$ i.e.\
\begin{equation}
	\tilde{\boldsymbol{z}}_{\text{rotated}}[k] = \boldsymbol{U}_{k}^{\dagger}\tilde{z}[k]\,.
\end{equation}

\subsection{Polarization angle}
\label{app:pol_angle}

In the most general case where the polarization modes are independent, we may write the GW transient signal $\tilde{\boldsymbol{s}}$ with each of the polarization components as follows
\begin{equation}
	\begin{split}
		\tilde{\boldsymbol{s}}_{\text{tensor}}
		&=
		\begin{bmatrix}
			\boldsymbol{f}_{+}(\psi) & \boldsymbol{f}_{\times}(\psi)
		\end{bmatrix}
		\begin{bmatrix}
			\tilde{h}_{+} \\
			\tilde{h}_{\times}
		\end{bmatrix}\\
		&=
		\begin{bmatrix}
			\boldsymbol{f}_{+}(\psi_{1}) & \boldsymbol{f}_{\times}(\psi_{1})
		\end{bmatrix}
		\boldsymbol{R}_{\text{tensor}}(\psi_{2})
		\begin{bmatrix}
			\tilde{h}_{+} \\
			\tilde{h}_{\times}
		\end{bmatrix}\\
		&=
		\begin{bmatrix}
			\boldsymbol{f}_{+}(\psi_{1}) & \boldsymbol{f}_{\times}(\psi_{1})
		\end{bmatrix}
		\left(
		\boldsymbol{R}_{\text{tensor}}(\psi_{2})
		\begin{bmatrix}
			\tilde{h}_{+} \\
			\tilde{h}_{\times}
		\end{bmatrix}
		\right)
	\end{split}
\end{equation}
where $\psi=\psi_{1}+\psi_{2}$ and
\begin{equation}
	\boldsymbol{R}_{\text{tensor}}(\psi)=
	\begin{bmatrix}
		\cos{2\psi} & -\sin{2\psi} \\
		\sin{2\psi} & \cos{2\psi}
	\end{bmatrix}
\end{equation}
is a rotational matrix that represents the rotation of coordinates around the GW-propagating axis. Similarly,
\begin{equation}
	\begin{split}
		\tilde{\boldsymbol{s}}_{\text{vector}}
		&=
		\begin{bmatrix}
			\boldsymbol{f}_{x}(\psi) & \boldsymbol{f}_{y}(\psi)
		\end{bmatrix}
		\begin{bmatrix}
			\tilde{h}_{x} \\
			\tilde{h}_{y}
		\end{bmatrix}\\
		&=
		\begin{bmatrix}
			\boldsymbol{f}_{x}(\psi_{1}) & \boldsymbol{f}_{y}(\psi_{1})
		\end{bmatrix}
		\boldsymbol{R}_{\text{vector}}(\psi_{2})
		\begin{bmatrix}
			\tilde{h}_{x} \\
			\tilde{h}_{y}
		\end{bmatrix}\\
		&=
		\begin{bmatrix}
			\boldsymbol{f}_{x}(\psi_{1}) & \boldsymbol{f}_{y}(\psi_{1})
		\end{bmatrix}
		\left(
		\boldsymbol{R}_{\text{vector}}(\psi_{2})
		\begin{bmatrix}
			\tilde{h}_{x} \\
			\tilde{h}_{y}
		\end{bmatrix}
		\right)
	\end{split}
\end{equation}
where $\psi=\psi_{1}+\psi_{2}$ and
\begin{equation}
	\boldsymbol{R}_{\text{vector}}(\psi)=
	\begin{bmatrix}
		\cos{\psi} & -\sin{\psi} \\
		\sin{\psi} & \cos{\psi}
	\end{bmatrix}\,.
\end{equation}
Since the scalar beam pattern function is itself independent of $\psi$, we do not show it here. We could observe that the signal could be equally well decribed as lying in the subspace spanned by $\{\boldsymbol{f}_{m}(\psi)\}$ for any $\psi$. This is due to the fact that rotation of the axes does not change the subspace spanned by $\{\boldsymbol{f}_{m}\}$. Therefore, the null operator construction in Eq.~\eqref{eq:P_def} is independent of $\psi$.

\section{Calibration errors}
\label{app:cal_error}
Following the similar notations used in Ref.~\cite{Vitale_2012}, given the measured strain output $\tilde{d}_{m}(f)$ and the exact strain output $\tilde{d}_{e}(f)$, the errors in amplitude and phase could be accounted by introducing a function $K(f)$
\begin{equation}
	\tilde{d}_{m}(f) = K(f)\tilde{d}_{e}(f)
\end{equation}
where
\begin{equation}
	K(f) = \left(1+\frac{\delta A(f)}{A(f)}\right)e^{i\delta\phi(f)}\,.
\end{equation}
Then the noise PSD with calibration errors is related to the noise PSD without calibration errors by
\begin{equation}
	S_{m}(f) = \left(1+\frac{\delta A(f)}{A(f)}\right)^{2}S_{e}(f)\,.
\end{equation}

Now we consider the effect of calibration errors in the construction of the null stream. For generality, we write the derivations in the continuous version here, but it is trivial to convert them to the discrete version. After including the calibration errors, the single-detector observation model writes
\begin{equation}
	K(f)\tilde{d}_{e}(f) = K(f)\left(\mathbf{F}\tilde{\mathbf{h}}(f)\right) + K(f)\tilde{n}_{e}(f)\,.
	\label{eq:obs_model_cal}
\end{equation}
The trivial way to recover the observation model with exact measurements in Eq.~\eqref{eq:obs_model_0} is to divide Eq.~\eqref{eq:obs_model_cal} by $K(f)$, but $K(f)$ is not known exactly. In the construction of null stream, we first whiten the data in Eq.~\eqref{eq:d_w}, however
\begin{equation}
	\begin{split}
	&\tilde{d}_{w}(f) \\
	&= \frac{\left(1+\frac{\delta A(f)}{A(f)}\right)e^{i\delta\phi(f)}\tilde{d}_{e}(f)}{\sqrt{\frac{1}{2\Delta f}\left(1+\frac{\delta A(f)}{A(f)}\right)^{2}S_{e}(f)}}\\
	&=\frac{e^{i\delta\phi(f)}\tilde{d}_{e}(f)}{\sqrt{\frac{1}{2\Delta f}S_{e}(f)}}
	\end{split}
\end{equation}
, and therefore the amplitude error is canceled, and we can conclude only the phase error affects the construction of null stream. The phase error is removed by multiplying the whitened data by $e^{-i\delta\phi(f)}$. Since the phase error is not known exactly, we use the conventional cubic spline model \cite{calnote} to model the phase error. Nodal points are equally spaced in $\log{f}$, and a Gaussian prior is placed at each node $j$ i.e.
\begin{equation}
	p(\delta\phi_{j}) = N(\mu_{j},\sigma_{j}^{2})
\end{equation}
where $\mu_{j}$ and $\sigma_{j}^{2}$ are mean and variance of phase error at node $j$ which characterize the expected phase error. The integral evaluating the model evidence in Eq.~\eqref{eq:evidence} is therefore extended to also marginalize over the phase errors. The calibration error characterization is released as the calibration envelope files by the LIGO Scientific Collaboration and Virgo Collaboration \cite{calfile}.
\clearpage

\bibliography{main}

\end{document}